\documentclass[aps,prb,twocolumn,superscriptaddress,eqsecnum,showpacs,showkeys,amsmath,amssymb]{revtex4}

\usepackage{amsfonts}
\usepackage{amssymb,amsmath}
\usepackage{graphicx}
\usepackage{dcolumn}
\usepackage{bm}
\usepackage{color}

\begin{document}

\title{Exact solution of a spin-$1/2$ $XX$ chain with three-site interactions in a random transverse field:
       Influence of randomness on quantum phase transition}

\author{Volodymyr Derzhko}
\affiliation{Institute of Theoretical Physics,
             University of Wroc\l aw,
             pl. Maksa Borna 9, 50-204 Wroc\l aw, Poland}

\author{Oleg Derzhko}
\affiliation{Institute for Condensed Matter Physics,
             National Academy of Sciences of Ukraine,
             1 Svientsitskii Street, L'viv-11, 79011, Ukraine}
\affiliation{Department for Theoretical Physics,
             Ivan Franko National University of L'viv,
             12 Drahomanov Street, L'viv-5, 79005, Ukraine}
\affiliation{Institut f\"ur Theoretische Physik,
             Universit\"at Magdeburg,
             P.O. Box 4120, 39016 Magdeburg, Germany}

\author{Johannes Richter}
\affiliation{Institut f\"ur Theoretische Physik,
             Universit\"at Magdeburg,
             P.O. Box 4120, 39016 Magdeburg, Germany}

\date{\today}

\begin{abstract}
We present exact results for the ground-state and thermodynamic properties 
of the spin-1/2 $XX$ chain with three-site interactions in a random (Lorentzian) transverse field. 
We discuss the influence of randomness on the quantum critical behavior 
known to be present in the nonrandom model. 
We find that at zero temperature the characteristic features of the quantum phase transition, 
such as kinks in the magnetization versus field curve, 
are smeared out by randomness.
However, 
at low but finite temperatures signatures of the quantum critical behavior are preserved 
if the randomness is not too large. 
Even the quantum critical region may be slightly enlarged for very weak randomness.
In addition to the exact results for Lorentzian randomness 
we present a more general discussion of an arbitrarily random transverse magnetic field  
based on the inspection of the moments of the density of states.
\end{abstract}



\maketitle

\section{Introductory remarks}
\label{sec1}
\setcounter{equation}{0}

In recent years the theory of quantum phase transitions 
has been in the focus of very active research.\cite{qpt1,qpt2,qpt2a}
The quantum phase transitions take place at zero temperature by changing a control parameter 
and emerge as a result of competing different ground-state phases.
Importantly, 
quantum phase transitions can influence the behavior of systems 
over a wide range of the phase diagram at nonzero (sometimes quite large) temperatures.
Exactly solvable quantum models exhibiting a quantum phase transition are notoriously rare. 
A well-known example of a solvable model is the spin-1/2 Ising chain in a transverse field,
where a zero-temperature transition 
from the ordered quantum Ising phase (small transverse fields) 
to the disordered quantum paramagnetic phase (large transverse fields) 
takes place. 
This model is often used for illustration of basic concepts 
in the quantum phase transition theory.\cite{pfeuty,luck,qpt3,qpt2,igloi,pre}
In general, 
spin-1/2 $XY$ chains\cite{lieb} provide an excellent ground for various statistical mechanics studies
since in many cases the calculations can be performed without any approximation.
Moreover, 
there are some real-life compounds 
which can be viewed as realizations of one-dimensional spin-1/2 $XY$ models.\cite{r1,r2,r3,r4}

Quite recently, two other classes of solvable models have been found, 
namely 
a two-dimensional Kitaev model\cite{kitaev} 
and  
a spin-1/2 $XY$ chain with multi-site interactions\cite{gottlieb_drd,titvinidze,lou}
(see also Ref.~\onlinecite{suzuki}).
The model belonging to the latter class is of interest in this paper.
This model has an essentially richer ground-state phase diagram 
as the standard one-dimensional spin-1/2 $XY$ model. 
In particular,   
it may exhibit several gapless spin-liquid phases 
and quantum phase transitions between them.\cite{titvinidze,lou}

On the other hand, 
quantum models with random Hamiltonian parameters present another class of models 
for which an exact solution cannot be found easily.
A solvable model with diagonal Lorentzian disorder was introduced by  Lloyd.\cite{lloyd}
Later on Lloyd's idea was used to study random spin-1/2 $XX$ chains.\cite{nishimori}
Also an extension to correlated off-diagonal Lorentzian disorder and its application to spin-1/2 $XX$ chains 
was considered, see Refs.~\onlinecite{ext_lloyd,dr}.

Naturally, 
the investigation of quantum phase transitions in systems with randomness is a challenging task. 
The random transverse-field Ising spin chain is known as a tractable model 
to study effects of quenched randomness on critical behavior.\cite{ising_chain}
Within the context of random quantum systems exactly solvable models may play an important role.
Merging together 
the above mentioned solvable quantum spin models with three-site interactions 
and 
Lloyd's model of disorder
we present here an exact analysis of a specific random quantum spin model.
In particular, 
the influence of randomness on the quantum phase transition
inherent in the nonrandom spin model 
can be studied.
In the presence of randomness the quantum phase transition becomes a crossover. 
Our solution presented below is based on the Jordan-Wigner transformation
of the spin Hamiltonian to the Hamiltonian of a tight-binding chain of spinless fermions
with nearest-neighbor and next-nearest-neighbor hoppings and random (Lorentzian) on-site energy.
Next-nearest-neighbor hopping is a new feature emerging owing to three-site interactions
which makes further calculations more involved.
We introduce Green functions and find exactly the random-averaged Green functions
which yield the random-averaged density of states.
We use the obtained density of states to discuss some ground-state and finite-temperature properties of the spin model.
Although the random-averaged density of states can be obtained only for a specific probability distribution,
the moments of the density of states can be obtained for an arbitrary inhomogeneous spin chain. 
These quantities can illustrate some general effects on the properties of the quantum spin chain caused by inhomogeneity
and yields thermodynamic quantities in the high-temperature limit. 
Our exact results allow to illustrate effect of randomness on a quantum phase transition.
In particular we discuss how a quantum critical region may be modified owing to randomness.

The paper is organized as follows.
In the next sections we define the spin model under consideration (Sec.~\ref{sec2a})
and calculate the random-averaged density of states which yields thermodynamic quantities (Sec.~\ref{sec2b}).
Then, in Sec.~\ref{sec3},
we discuss some properties of the spin model at zero and nonzero temperatures.
We illustrate the effect of the introduced disorder 
for the transverse magnetization, specific heat, and static transverse susceptibility,
and put our discussion in a general context of a theory of quantum phase transitions with randomness.
In Sec.~\ref{sec4} we discuss some global properties of the density of states
for an arbitrary inhomogeneous spin-$1/2$ transverse $XX$ chain with three-site interactions.
Finally, in Sec.~\ref{sec5}, we summarize our findings.

\section{The model}
\label{sec2a}
\setcounter{equation}{0}

To be specific, 
we consider a linear chain of $N$ spins with spin quantum number $s=1/2$.
Each spin interacts with spins on nearest-neighboring sites and on next-nearest-neighboring sites.
Moreover, 
all spins interact with an external magnetic field which acquires a random value on each site.
The Hamiltonian of the model reads
\begin{eqnarray}
\label{2.01}
H=\sum_{n}
\left[J\left(s_n^xs_{n+1}^x+s_n^ys_{n+1}^y\right)
\right.
\nonumber\\
\left.
+K\left(s_n^xs^z_{n+1}s_{n+2}^x+s_n^ys^z_{n+1}s_{n+2}^y\right)\right]
+\sum_{n}\Omega_ns_n^z,
\end{eqnarray}
where periodic boundary conditions are implied for convenience.
Here $J$ and $K$ are the two-site isotropic $XY$ (i.e., $XX$) interaction and the three-site $XZX+YZY$ interaction, 
respectively,
and $\Omega_n$ is the transverse field on the site $n$.
Although, exact solvability is the main motivation to consider the
three-site interactions, we note that 
 Hamiltonians similar to Eq. (\ref{2.01}) may be generated in optical lattices.\cite{opt_lat}

The on-site transverse fields are assumed to be independent random variables
each with the Lorentzian probability distribution
\begin{eqnarray}
\label{2.02}
p(\Omega_n)=\frac{1}{\pi}\frac{\Gamma}{(\Omega_n-\Omega_0)^2+\Gamma^2},
\end{eqnarray}
where $\Omega_0$ is the mean value and $\Gamma$ controls the strength of disorder.
We are interested in (random-averaged) thermodynamic quantities of the spin model (\ref{2.01}), (\ref{2.02}).

As the first step in the calculation of thermodynamic quantities of the spin model 
we perform the Jordan-Wigner fermionization\cite{lieb}
to transform the Hamiltonian (\ref{2.01}) into a bilinear Fermi form
\begin{eqnarray}
\label{2.03}
H=\sum_{n}
\left[\frac{J}{2}\left(c_n^{\dagger}c_{n+1}+c_{n+1}^{\dagger}c_n\right)
\right.
\nonumber\\
\left.
-\frac{K}{4}\left(c_n^{\dagger}c_{n+2}+c_{n+2}^{\dagger}c_n\right)
+\Omega_n\left(c_n^{\dagger}c_n-\frac{1}{2}\right)\right]
\nonumber\\
=\sum_{n=1}^N\sum_{m=1}^Nc_n^{\dagger}A_{nm}c_m-\frac{1}{2}\sum_{n=1}^N\Omega_n.
\end{eqnarray}
As typical for the fermionic representation of the spin model 
the magnetic field (here its uniform part $\Omega_0$) plays the role of a chemical potential.
From Ref.~\onlinecite{lieb} we know
that the bilinear form in Eq. (\ref{2.03}) can be diagonalized.
After performing the linear canonical transformation
\begin{eqnarray}
\label{2.04}
\eta_\nu=\sum_{n=1}^Ng_{\nu n}c_n,
\;\;\;
\eta_\nu^{\dagger}=\sum_{n=1}^Ng_{\nu n}c_n^{\dagger},
\nonumber\\
\Lambda_\nu g_{\nu n}=\sum_{i=1}^Ng_{\nu i}A_{in},
\nonumber\\
\sum_{i=1}^Ng_{\nu i}g_{\mu i}=\delta_{\nu \mu},
\;\;\;
\sum_{\mu=1}^Ng_{\mu i}g_{\mu j}=\delta_{ij}
\end{eqnarray}
we find
\begin{eqnarray}
\label{2.05}
H=\sum_{\nu =1}^N\Lambda_\nu\left(\eta^{\dagger}_\nu\eta_\nu-\frac{1}{2}\right).
\end{eqnarray}
Although this can be done in principle,
to find $g_{\nu n}$ and $\Lambda_\nu$ is a complicated task in practice because of nonhomogeneous values of $\Omega_n$.

\begin{figure}
\begin{center}
\includegraphics[clip=on,width=7.5cm,angle=0]{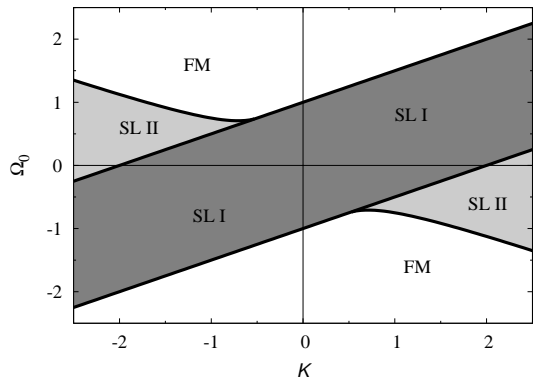}
\caption{The ground-state phase diagram of nonrandom model (\ref{2.01}) with $J=\pm 1$ and $\Omega_n=\Omega_0$ 
discussed earlier in Refs.~\onlinecite{titvinidze,kdsv}.
Dark gray region corresponds to the spin-liquid I phase (two Fermi points),
light gray regions correspond to the spin-liquid II phase (four Fermi points),
and
white regions correspond to the ferromagnetic phase.}
\label{fig00}
\end{center}
\end{figure}

Before we present the solution of the random model, 
for convenience we illustrate briefly the basic features of the nonrandom model,
i.e., $\Omega_n=\Omega_0$ is independent of the site index $n$,
see Refs.~\onlinecite{titvinidze,kdsv} and Fig.~\ref{fig00}.
The nonrandom model exhibits three phases.
In the $K$--$\Omega_0$ plane 
($\vert J\vert=1$),
the spin-liquid I phase occurs in the region 
$-1+K/2<\Omega_0<1+K/2$ (dark gray region in Fig.~\ref{fig00}),
the spin-liquid II phase occurs in the regions 
$K<-1/2$, $1+K/2<\Omega_0<-K/2-1/(4K)$
and
$K>1/2$, $-K/2-1/(4K)<\Omega_0<-1+K/2$ (light gray regions in Fig.~\ref{fig00}),
whereas the remaining regions (white in Fig.~\ref{fig00}) correspond to the ferromagnetic phase.
The two different spin-liquid phases correspond to gapless spinless-fermion systems having two or four Fermi points,
whereas the ferromagnetic phase corresponds to a gapped spinless-fermion system.
Although the choice of the order parameter
which in a transparent way would be associated with the modification of the Fermi-surface topology 
is still under debate,\cite{titvinidze,lou,kdsv,order_parameter}
there is no doubt that the different phases and the transitions between them may be identified 
looking at the behavior of the ground-state transverse magnetization $m_z$ as a function of $\Omega_0$ or $K$.
Several cusps in the magnetization curve indicate quantum phase transition points,
see curves for $\Gamma=0$ in Figs.~\ref{fig02} and \ref{fig03} below. 

\section{The averaged density of states and thermodynamic quantities}
\label{sec2b}
\setcounter{equation}{0}

The free-fermion representation of the spin model, 
Eq. (\ref{2.05}), 
immediately implies simple formulas for thermodynamic quantities, as
\begin{eqnarray}
\label{2.06}
f=-T\int{\rm{d}}\omega\rho(\omega)\ln\left(2\cosh\frac{\omega}{2T}\right),
\nonumber\\
\rho(\omega)=\frac{1}{N}\sum_{\nu=1}^N\delta(\omega-\Lambda_{\nu})
\end{eqnarray}
for the Helmholtz free energy (per site).
Here we have introduced the density of states $\rho(\omega)$.
The random-averaged Helmholtz free energy $\overline{f}$ is given by Eq. (\ref{2.06})
with the random-averaged density of states $\overline{\rho(\omega)}$,
where
$\overline{(\ldots)}
=\prod_{n=1}^N\int{\rm{d}}\Omega_np(\Omega_n)(\ldots)$.
Thus our task is to find $\overline{\rho(\omega)}$.

Using (\ref{2.06}), (\ref{2.05}), (\ref{2.04}) one can easily convince oneself that
\begin{eqnarray}
\label{2.07}
\rho(\omega)
=\mp\frac{1}{N\pi}\sum_{j=1}^N\Im G^{\mp}_{jj}(\omega\pm{\rm{i}}\epsilon),
\end{eqnarray}
where
\begin{eqnarray}
\label{2.08}
G_{nm}^{\mp}(t)=\mp{\rm{i}}\theta(\pm t)
\langle\{c_n(t),c_m^{\dagger}\}\rangle,
\nonumber\\
G_{nm}^{\mp}(t)
=\frac{1}{2\pi}\int_{-\infty}^{\infty}{\rm{d}}\omega
\exp(-{\rm{i}}\omega t)G_{nm}^{\mp}(\omega\pm{\rm{i}}\epsilon)
\end{eqnarray}
[$\theta(x)$ is the Heaviside step function]
are the retarded and advanced temperature double-time Green functions.\cite{zubarev,rickayzen,mahan}
On the other hand,
one easily finds the following set of equations for $G_{nm}^{\mp}(\omega\pm{\rm{i}}\epsilon)$
\begin{eqnarray}
\label{2.09}
\left(\omega\pm{\rm{i}}\epsilon-\Omega_n\right)G_{nm}^{\mp}(\omega\pm{\rm{i}}\epsilon)
\nonumber\\
-\frac{J}{2}\left[G_{n-1,m}^{\mp}(\omega\pm{\rm{i}}\epsilon)+G_{n+1,m}^{\mp}(\omega\pm{\rm{i}}\epsilon)\right]
\nonumber\\
+\frac{K}{4}\left[G_{n-2,m}^{\mp}(\omega\pm{\rm{i}}\epsilon)+G_{n+2,m}^{\mp}(\omega\pm{\rm{i}}\epsilon)\right]
=\delta_{nm}.
\end{eqnarray}
Because of nonhomogeneous values of $\Omega_n$ it is not possible to solve (\ref{2.09})
and to find the required diagonal Green functions $G_{nn}^{\mp}(\omega\pm{\rm{i}}\epsilon)$ 
which enter Eq. (\ref{2.07}).
However, it is well known\cite{lloyd}
that if $\Omega_n$ is a Lorentzian random variable (\ref{2.02})
the set of equations (\ref{2.09}) can be averaged over random realizations
leading to a set of equation for translational-invariant
random-averaged Green functions $\overline{G_{nm}^{\mp}(\omega)}$.
Supposing that $\Omega_n$ is a complex variable
and noticing that $G_{nm}^{-}(\omega+{\rm{i}}\epsilon)$ [$G_{nm}^{+}(\omega-{\rm{i}}\epsilon)$]
cannot have a pole in the lower [upper] half-plane of the complex variable $\Omega_n$
we perform the averaging with (\ref{2.02}) by means of contour integrals
closing the contours of integrations in the half-planes 
where the Green function has no poles.\cite{lloyd,nishimori,ext_lloyd,dr}
As a result we obtain
\begin{eqnarray}
\label{2.10}
\left(\omega\pm{\rm{i}}\Gamma-\Omega_0\right)\overline{G_{nm}^{\mp}(\omega)}
\nonumber\\
-\frac{J}{2}\left[\overline{G_{n-1,m}^{\mp}(\omega)}+\overline{G_{n+1,m}^{\mp}(\omega)}\right]
\nonumber\\
+\frac{K}{4}\left[\overline{G_{n-2,m}^{\mp}(\omega)}+\overline{G_{n+2,m}^{\mp}(\omega)}\right]
=\delta_{nm}.
\end{eqnarray}

\begin{widetext}

The set of equations (\ref{2.10}) possesses translational symmetry already
and therefore
\begin{eqnarray}
\label{2.11}
\overline{G_{nm}^{\mp}(\omega)}
=\frac{1}{2\pi}\int_{-\pi}^{\pi}{\rm{d}}\kappa
\frac{\exp\left[{\rm{i}}(n-m)\kappa\right]}{\omega-\Omega_0-J\cos\kappa+\frac{K}{2}\cos(2\kappa)\pm{\rm{i}}\Gamma}.
\end{eqnarray}
To evaluate the integral in (\ref{2.11}) we introduce a new variable $z=\exp({\rm{i}}\kappa)$.
Then Eq. (\ref{2.11}) becomes
\begin{eqnarray}
\label{2.12}
\overline{G_{nm}^{\mp}(\omega)}
=\frac{1}{2\pi{\rm{i}}}
\oint{\rm{d}}z
\frac{z^{n-m+1}}{\frac{K}{4}(z^4+1)-\frac{J}{2}z(z^2+1)+(\omega-\Omega_0\pm{\rm{i}}\Gamma)z^2},
\end{eqnarray}
where the contour of integration runs counterclockwise along the unit circle in the complex plane $z$.

\end{widetext}

The calculation of (\ref{2.12}) is simple if either $K=0$ or $J=0$
yielding
$\overline{G_{nn}^{\mp}(\omega)}
=1/\sqrt{(\omega-\Omega_0\pm{\rm{i}}\Gamma)^2-J^2}$
(see Ref.~\onlinecite{nishimori})
or
$\overline{G_{nn}^{\mp}(\omega)}
=1/\sqrt{(\omega-\Omega_0\pm{\rm{i}}\Gamma)^2-K^2/4}$.
For arbitrary values of the interaction constants, $0<\vert K/J\vert<\infty$,
we have to solve the 4th order algebraic equation
\begin{eqnarray}
\label{2.13}
z^4-\frac{2J}{K}z^3+\frac{4}{K}\left(\omega-\Omega_0\pm{\rm{i}}\Gamma\right)z^2-\frac{2J}{K}z+1=0
\end{eqnarray}
which is a quasi-symmetric one
(i.e., of the form $a_0z^4+a_1z^3+a_2z^2+a_1mz+a_0m^2=0$ with $m=1$).
Dividing Eq. (\ref{2.13}) by $z^2$ and using the variable change $y=z+1/z$
we immediately find
\begin{eqnarray}
\label{2.14}
y_{\pm}=\frac{J}{K}\pm g,
\nonumber\\
g=\sqrt{\frac{J^2}{K^2}-\frac{4}{K}\left(\omega-\Omega_0\pm{\rm{i}}\Gamma\right)+2}.
\end{eqnarray}
As a result,
\begin{eqnarray}
\label{2.15}
z_{\pm}=\frac{y\pm \sqrt{y^2-4}}{2}, 
\;\;\;
z_+z_-=1,
\end{eqnarray}
where $y$ is either $y_+$ or $y_-$.
Let us denote the roots of Eq. (\ref{2.13}),
which are given in Eqs. (\ref{2.15}) and (\ref{2.14}),
by $z_1$, $z_2$, $z_3$, $z_4$,
$\vert z_1\vert\le\vert z_2\vert\le\vert z_3\vert\le\vert z_4\vert$,
see Appendix \ref{a}.
Only two roots are inside the unit circle $\vert z\vert<1$ resulting in 
\begin{eqnarray}
\label{2.16}
\overline{G_{nn}^{\mp}(\omega)}
=
\frac{4}{K}
\left[
\frac{z_1}{\left(z_1-z_2\right)\left(z_1-z_3\right)\left(z_1-z_4\right)}
\right.
\nonumber\\
\left.
+
\frac{z_2}{\left(z_2-z_1\right)\left(z_2-z_3\right)\left(z_2-z_4\right)}
\right].
\end{eqnarray}
Then the density of states $\overline{\rho(\omega)}$ is calculated according to Eq. (\ref{2.07}).
The described scheme for $\Gamma=0$ 
reproduces the density of states of the nonrandom model reported in Ref.~\onlinecite{titvinidze}.

\begin{figure}
\begin{center}
\includegraphics[clip=on,width=7.5cm,angle=0]{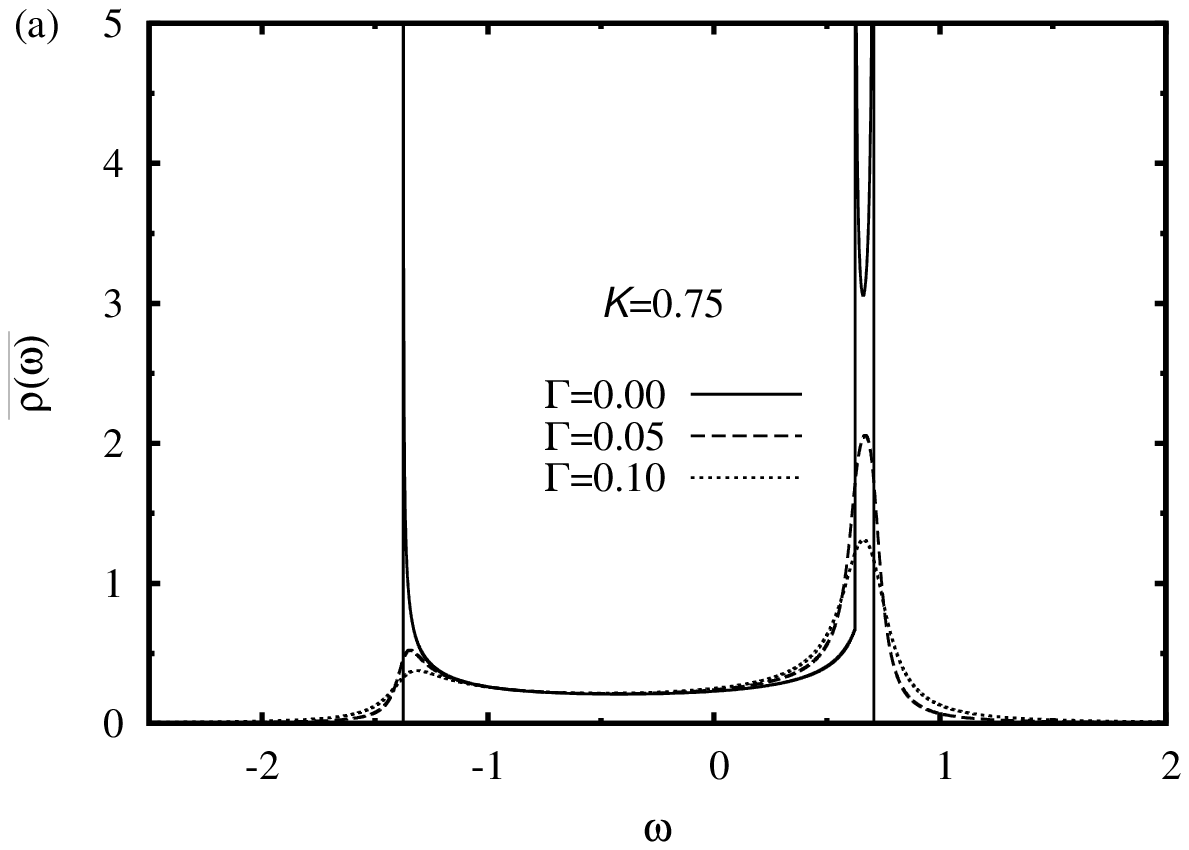}
\includegraphics[clip=on,width=7.5cm,angle=0]{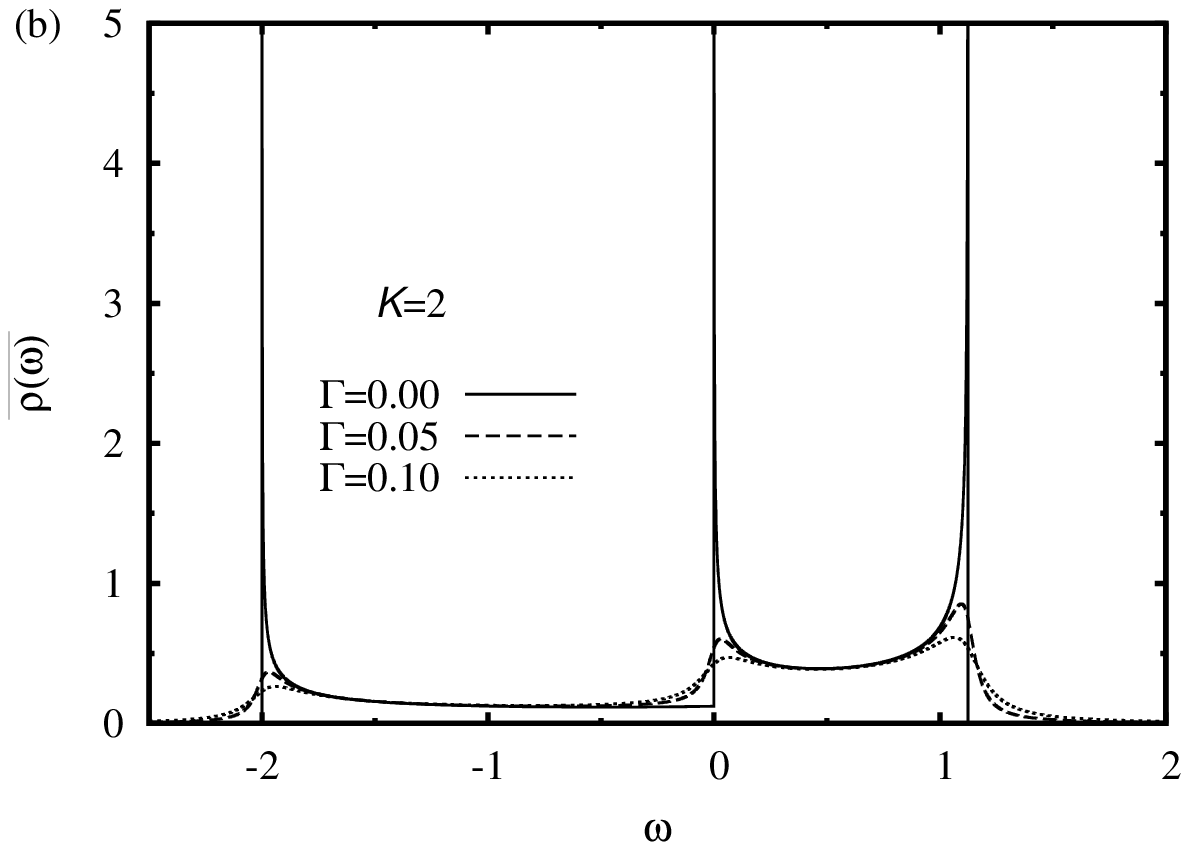}
\caption{The random-averaged density of states $\overline{\rho(\omega)}$
for the spin model (\ref{2.01}), (\ref{2.02}) with $J=1$,
$K=0.75$ (a), $K=2$ (b),
$\Omega_0=0$,
and
$\Gamma=0$ (solid), $\Gamma=0.05$ (dashed), $\Gamma=0.1$ (dotted).}
\label{fig01}
\end{center}
\end{figure}
Our results for $\overline{\rho(\omega)}$ for typical sets of parameters are shown in Fig.~\ref{fig01}.
We put $\Omega_0= 0$, 
since a nonzero $\Omega_0$ leads only to a trivial shift along the $\omega$-axis,
see, e.g., Eq. (\ref{2.11}). 
At the van Hove singularities present for the nonrandom model in Fig.~\ref{fig01} 
the density of states exhibits the typical one-dimensional inverse square-root singularity.
For parameters  $\Omega_0$ or $K$ where a quantum phase transition occurs, 
a van Hove singularity is located at $\omega=\Omega_0$
(as in Fig.~\ref{fig01}b, where it is at $\omega=\Omega_0=0$).
The middle peak in the density of states shown in Fig.~\ref{fig01}
(it appears if $\vert K\vert>1/2$, see Fig.~\ref{fig00})
indicates a quantum phase transition between the spin-liquid I and spin-liquid II phases.
Small randomness leads mainly to rounding of the van Hove singularities 
and to the appearance of tails in the density of states 
above and below the band edges of the nonrandom model.
With increasing  of $\Gamma$ the density of states becomes more and more smeared out, 
i.e., the fingerprints of the van Hove singularities disappear and the tails increase.
These changes of the density of states due to randomness 
influence the ground-state and finite-temperature properties to be discussed in the next section.

Finally, it is important to note the following.
A ground state of the nonrandom spin model (\ref{2.01}) 
is generally speaking a product of one-particle (more precisely, one-fermion) states, which are plane waves.
While for $\Gamma=0$ the one-particle states within the band(s) are extended states with infinite localization length,
for any small (diagonal) Lorentzian disorder, i.e., for $\Gamma>0$, 
{\it{all}} one-particle states of the Hamiltonian (\ref{2.05}) become localized, 
see, e.g., Ref.~\onlinecite{thouless}.
As a result,
the nature of the quantum phases of the nonrandom model changes 
and the quantum phase transition inherent in the nonrandom model becomes a crossover.\cite{others}

\section{Ground-state and thermodynamic properties}
\label{sec3}
\setcounter{equation}{0}

The obtained (random-averaged) density of states permits to examine various quantities 
characterizing behavior of the spin model at zero and nonzero temperatures, see Eq. (\ref{2.06}).
For the ground-state energy, the entropy, and the specific heat we have
\begin{eqnarray}
\label{3.01}
\overline{e_0}
=
\int{\rm{d}}\omega\overline{\rho(\omega)}\frac{\vert \omega\vert}{2},
\end{eqnarray}
\begin{eqnarray}
\label{3.02}
\overline{s}
=
\int{\rm{d}}\omega\overline{\rho(\omega)}
\left[\ln\left(2\cosh\frac{\omega}{2T}\right)
-\frac{\omega}{2T}\tanh\frac{\omega}{2T}\right],
\end{eqnarray}
\begin{eqnarray}
\label{3.03}
\overline{c}
=
\int{\rm{d}}\omega\overline{\rho(\omega)}
\left(\frac{\frac{\omega}{2T}}{\cosh\frac{\omega}{2T}}\right)^2,
\end{eqnarray}
respectively.
Next,
for the transverse magnetization and the static transverse susceptibility
we have
\begin{eqnarray}
\label{3.04}
\overline{m_z}
=\frac{\partial \overline{f}}{\partial\Omega_0}
=
-\frac{1}{2}\int{\rm{d}}\omega\overline{\rho(\omega)}\tanh\frac{\omega}{2T},
\end{eqnarray}
\begin{eqnarray}
\label{3.05}
\overline{\chi_{zz}}
=\frac{\partial \overline{m_z}}{\partial\Omega_0}
=
-\frac{1}{4T}\int{\rm{d}}\omega\overline{\rho(\omega)}
\frac{1}{\cosh^2\frac{\omega}{2T}},
\end{eqnarray}
respectively.

\begin{figure}
\includegraphics[clip=on,width=7.5cm,angle=0]{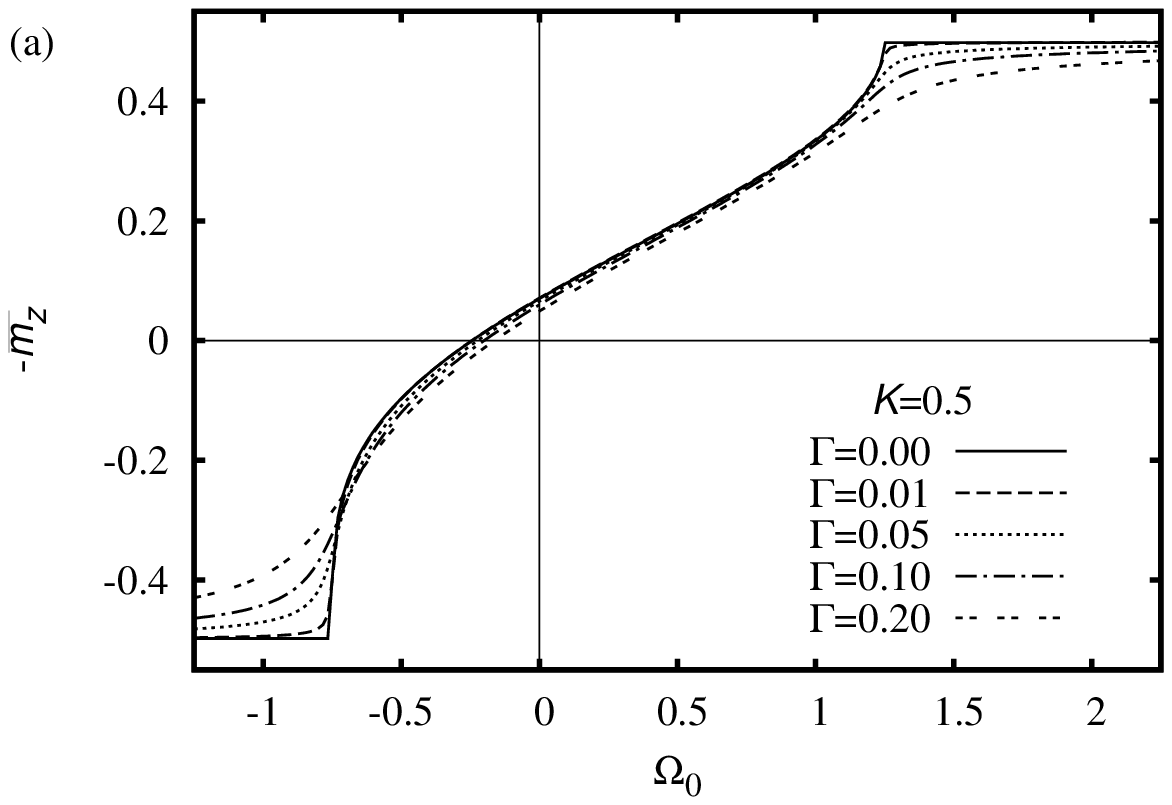}
\includegraphics[clip=on,width=7.5cm,angle=0]{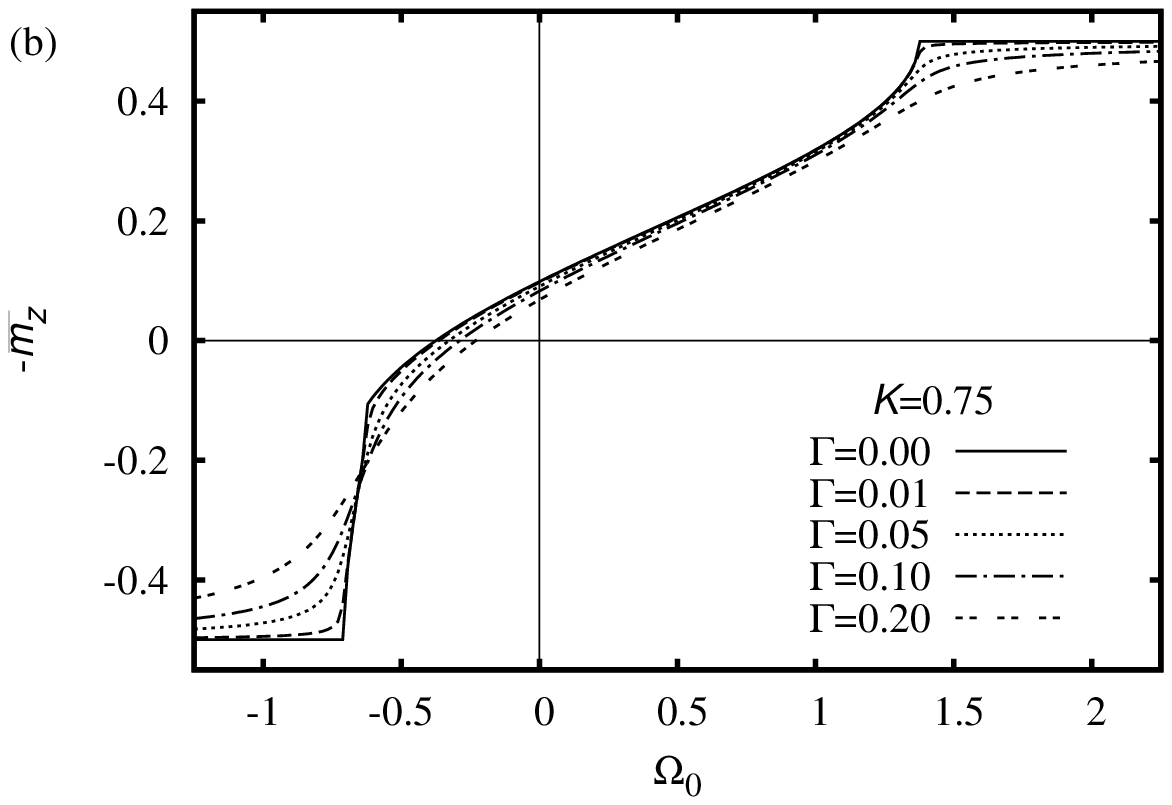}
\includegraphics[clip=on,width=7.5cm,angle=0]{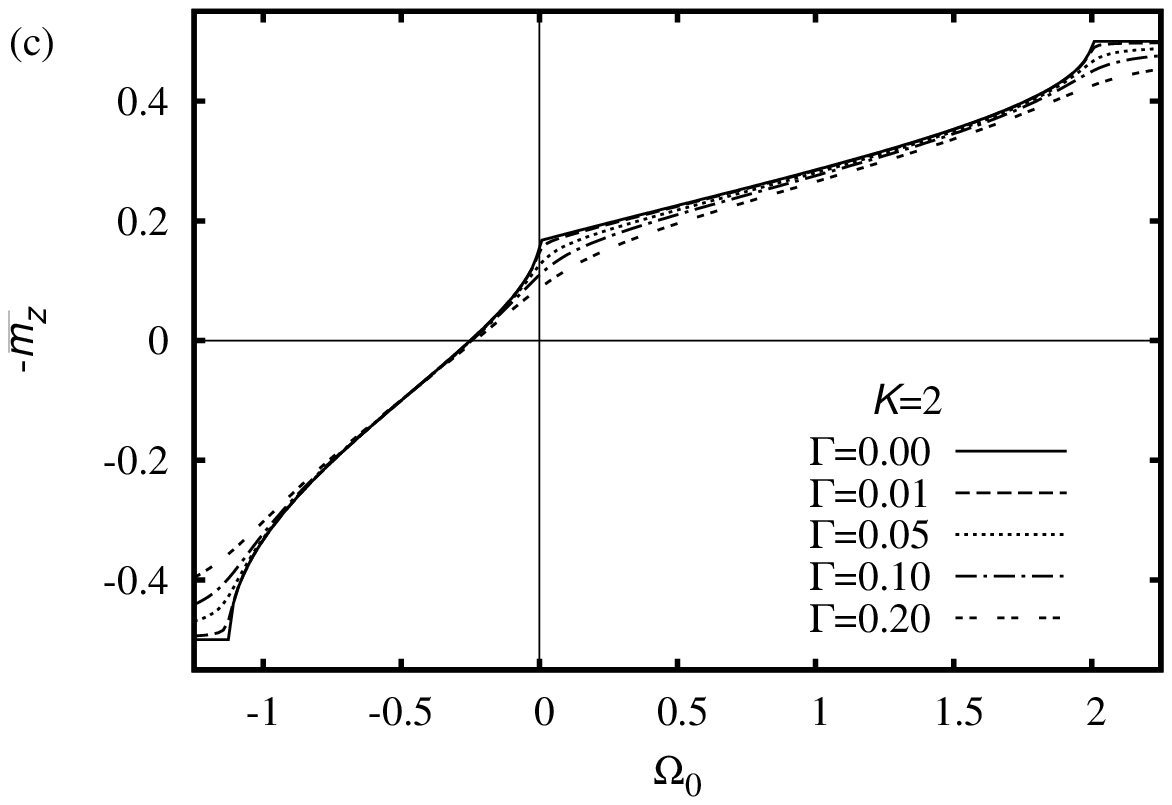}
\caption{Ground-state transverse magnetization $-\overline{m_z}$ (\ref{3.04}) versus $\Omega_0$
for the spin model (\ref{2.01}), (\ref{2.02}) with $J=1$,
$K=0.5$ (a), $K=0.75$ (b), $K=2$ (c) 
for $\Gamma=0,\ldots,0.2$.
The quantum phase transitions occurring in the nonrandom model 
at $\Omega_0=-0.75$, $\Omega_0=1.25$ (a), 
at $\Omega_0\approx -0.708$, $\Omega_0=-0.625$, $\Omega_0=1.375$ (b), 
and 
at $\Omega_0=-1.125$, $\Omega_0=0$, $\Omega_0=2$ (c)
are signaled by kinks in the magnetization.}
\label{fig02}
\end{figure}
\begin{figure}
\includegraphics[clip=on,width=7.5cm,angle=0]{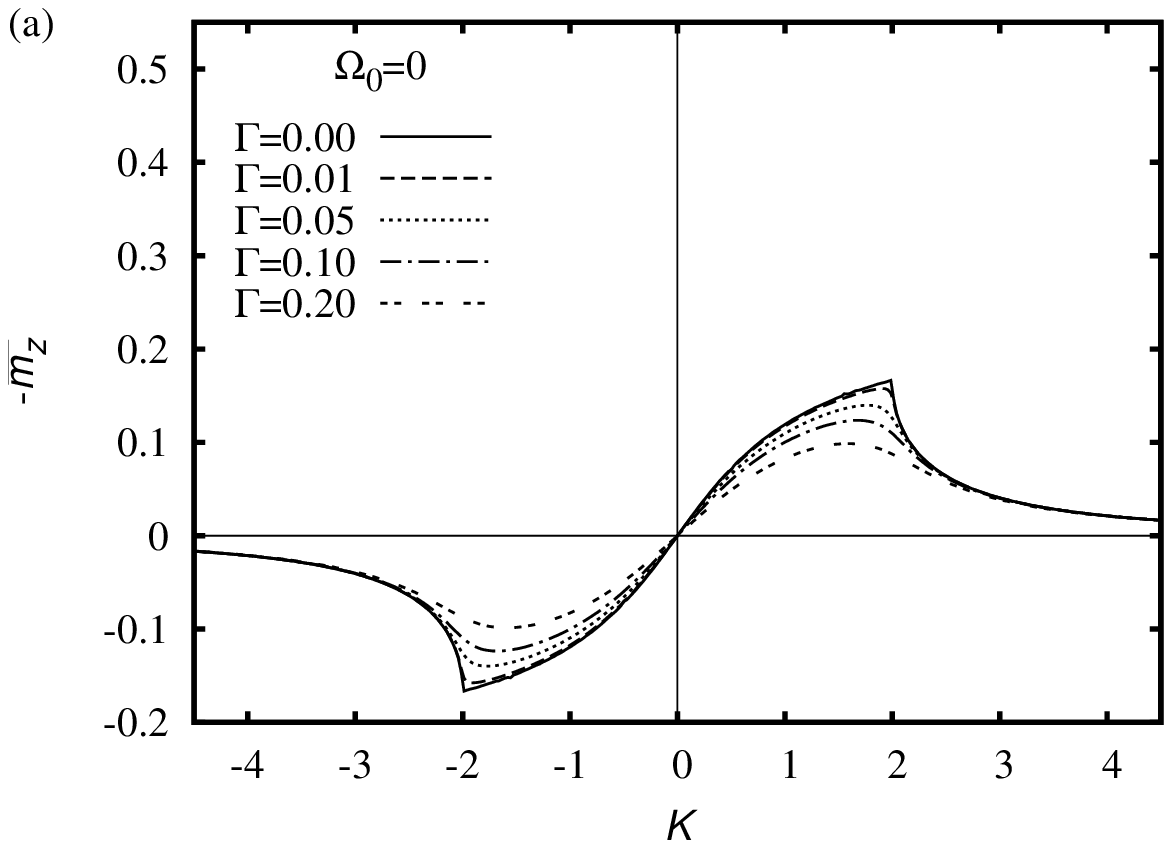}
\includegraphics[clip=on,width=7.5cm,angle=0]{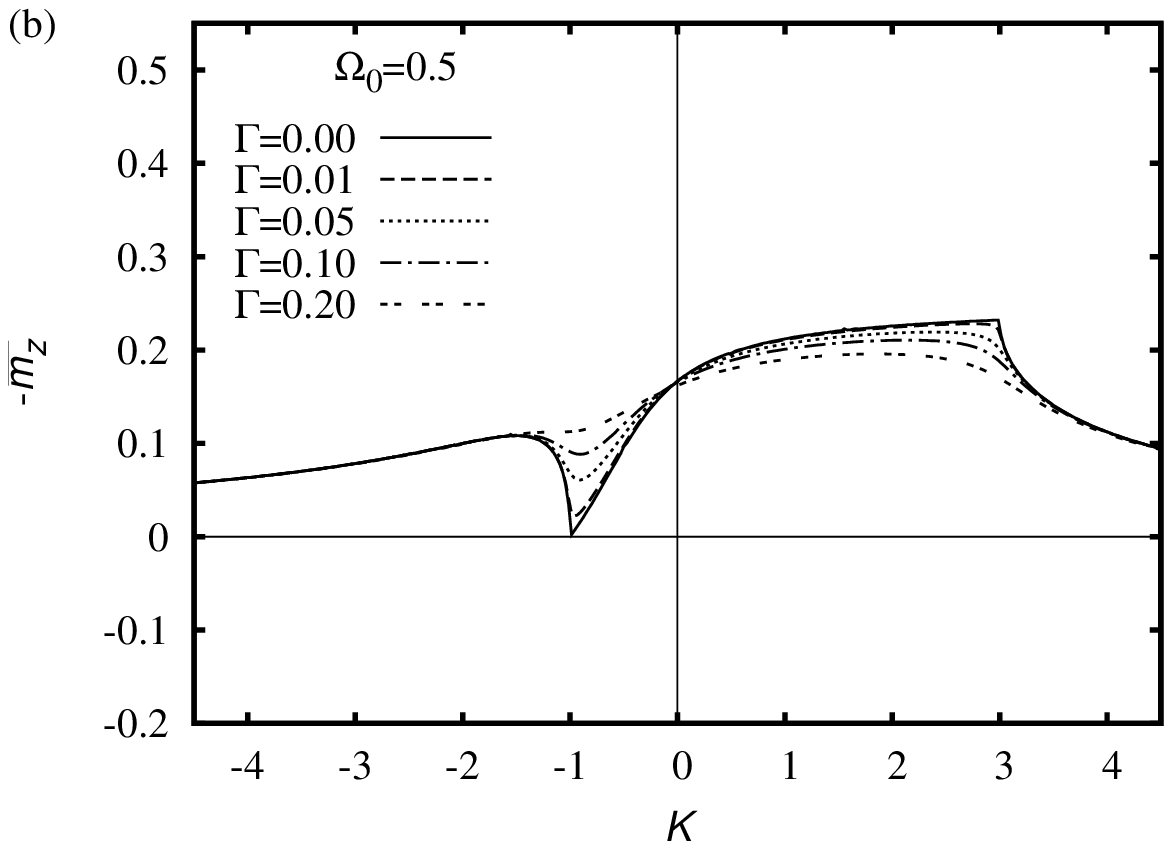}
\includegraphics[clip=on,width=7.5cm,angle=0]{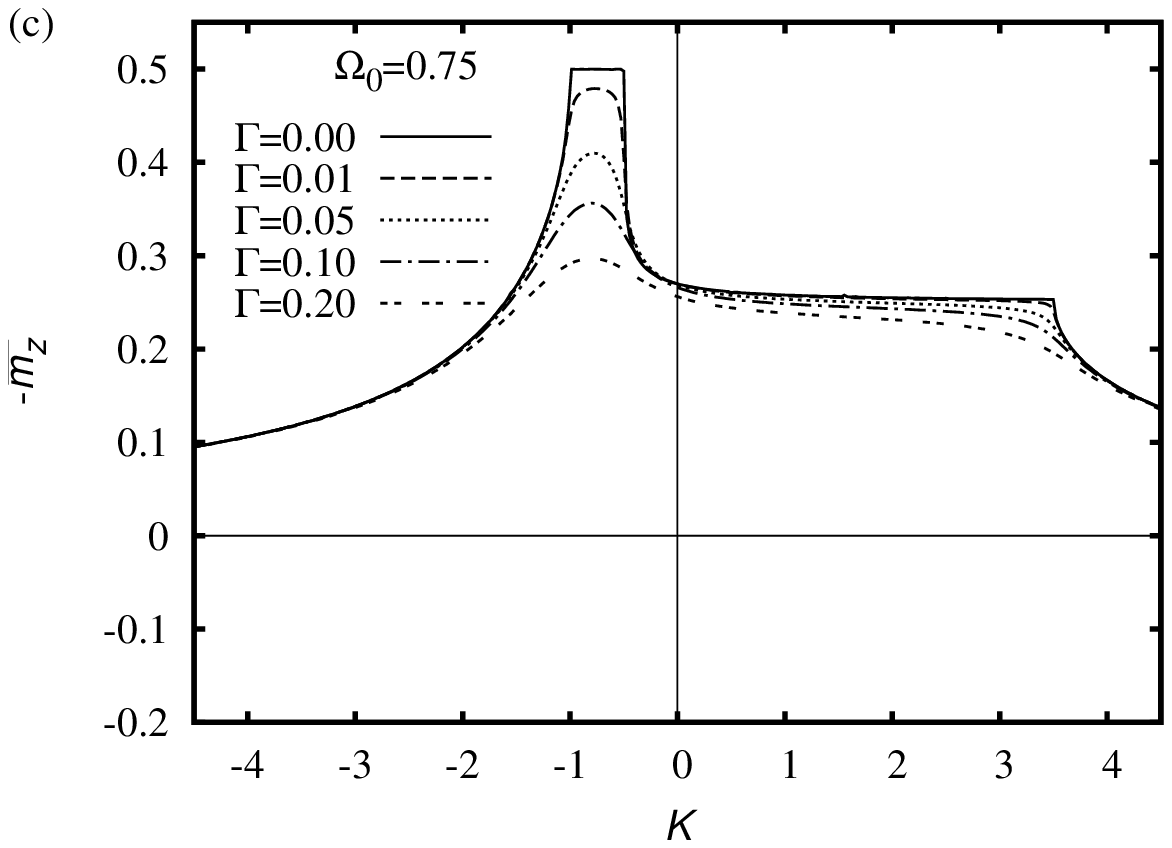}
\caption{Ground-state transverse magnetization $-\overline{m_z}$ (\ref{3.04}) versus $K$
for the spin model (\ref{2.01}), (\ref{2.02}) with $J=1$,
$\Omega_0=0$ (a), $\Omega_0=0.5$ (b), $\Omega_0=0.75$ (c)
for $\Gamma=0,\ldots,0.2$.
The quantum phase transitions occurring in the nonrandom model 
at $K=-2$, $K=2$ (a), 
at $K=-1$, $K=3$ (b),
and 
at $K=-1$, $K=-0.5$, $K=3.5$ (c) 
are signaled by kinks in the magnetization.}
\label{fig03}
\end{figure}
In Figs.~\ref{fig02} and \ref{fig03} we report some results for the ground-state transverse magnetization.
In the ground state of the nonrandom spin system 
the model can be in three different phases (spin-liquid I, II, and ferromagnetic), 
see Refs.~\onlinecite{titvinidze,kdsv} and the  discussion in Sec. \ref{sec2a}. 
The transitions between them 
can clearly be detected by the cusps in the magnetization curves in Figs.~\ref{fig02} and \ref{fig03}.

Let us briefly discuss a prominent feature of the magnetization curve, 
namely the steep part in the curve near saturation 
($\Omega_0\approx -0.708$)
seen in Fig.~\ref{fig02}b for $\Gamma=0$.
This jump-like behavior resembles the magnetization jumps 
observed in frustrated quantum antiferromagnets.\cite{loc_mag}
The corresponding density of states, 
see Fig.~\ref{fig01}a,
shows a narrow upper band, 
present for $K > 1/2$. 
The two singularities defining the band edges approach each other if $K \to 1/2$, 
i.e., the upper band becomes flat. 
However, by contrast to flat bands discussed in Ref.~\onlinecite{loc_mag} 
the number of states in the narrow upper band of our model decreases with decreasing of band width.
As a result the middle cusp in the magnetization curve, 
see Fig.~\ref{fig02}b, 
related to the middle singularity in the density of states,
see Fig.~\ref{fig01}a, 
moves to the left cusp this way yielding the steep part before saturation 
seen in Fig.~\ref{fig02}b. 
The slope of that part of the magnetization curve increases if $K \to 1/2$, 
however, at the same time its height decreases and vanishes finally at $K=1/2$, 
where the magnetization approaches the saturation continuously with an infinite slope,
see  Fig.~\ref{fig02}a.
Moreover,
$1/2-m_z\propto (\Omega_0+0.75)^{\varepsilon}$ if $\Omega_0+0.75\to +0$
with $\varepsilon=1/4$ instead of the usual value $\varepsilon=1/2$, see Ref.~\onlinecite{qpt2}.

The effect of randomness on the magnetization $m_z$ is similar to that of a finite temperature.
For small randomness the cusps in the $\overline{m_z}$-curves, 
which indicate boundaries of different ground-state phases, 
become rounded, 
indicating that a quantum phase transition present at $\Gamma=0$ 
transforms into a crossover at $\Gamma>0$.
Although even small randomness is sufficient to erode the boundaries between different ground-state phases 
by a noticeably rounding of the cusps of $\overline{m_z}$,  
it may have almost no influence on $\overline{m_z}$ for parameter values corresponding to the spin-liquid phases,
see Figs.~\ref{fig02} and \ref{fig03}.
Other peculiarities of the nonrandom model,  
namely nonzero magnetization at $\Omega_0=0$, zero magnetization at nonzero $\Omega_0$,
as well as saturated magnetization for $\vert\Omega_0\vert <\vert J\vert$,
see Fig.~\ref{fig03},
become less pronounced as the strength of disorder increases.

\begin{figure}
\includegraphics[clip=on,width=7.5cm,angle=0]{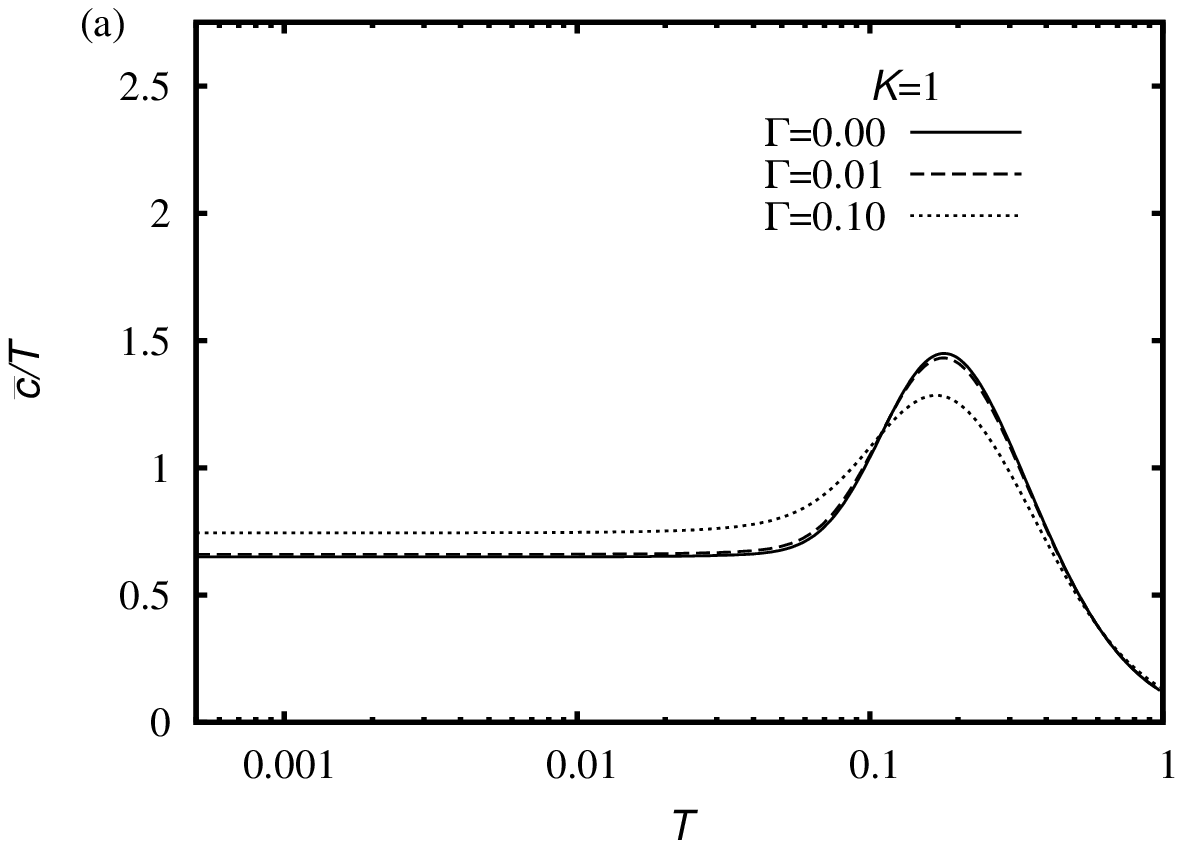}
\includegraphics[clip=on,width=7.5cm,angle=0]{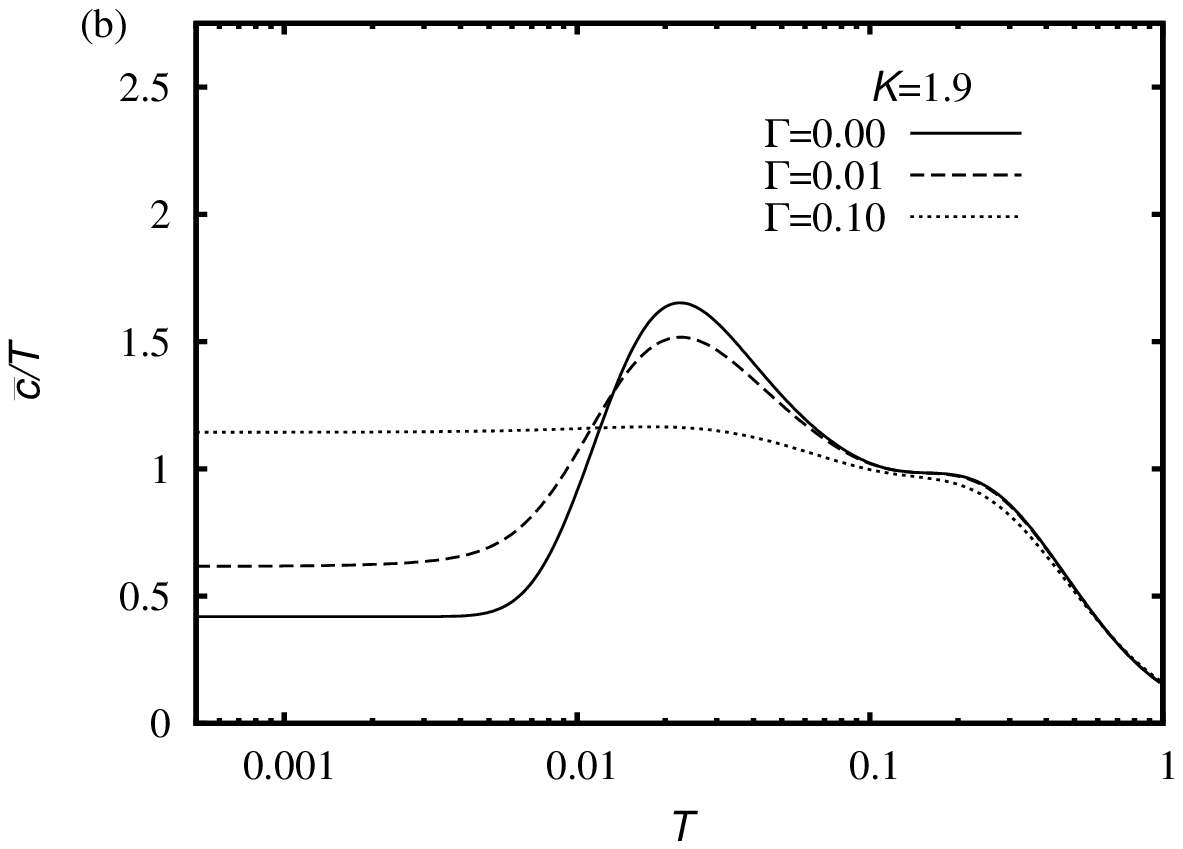}
\includegraphics[clip=on,width=7.5cm,angle=0]{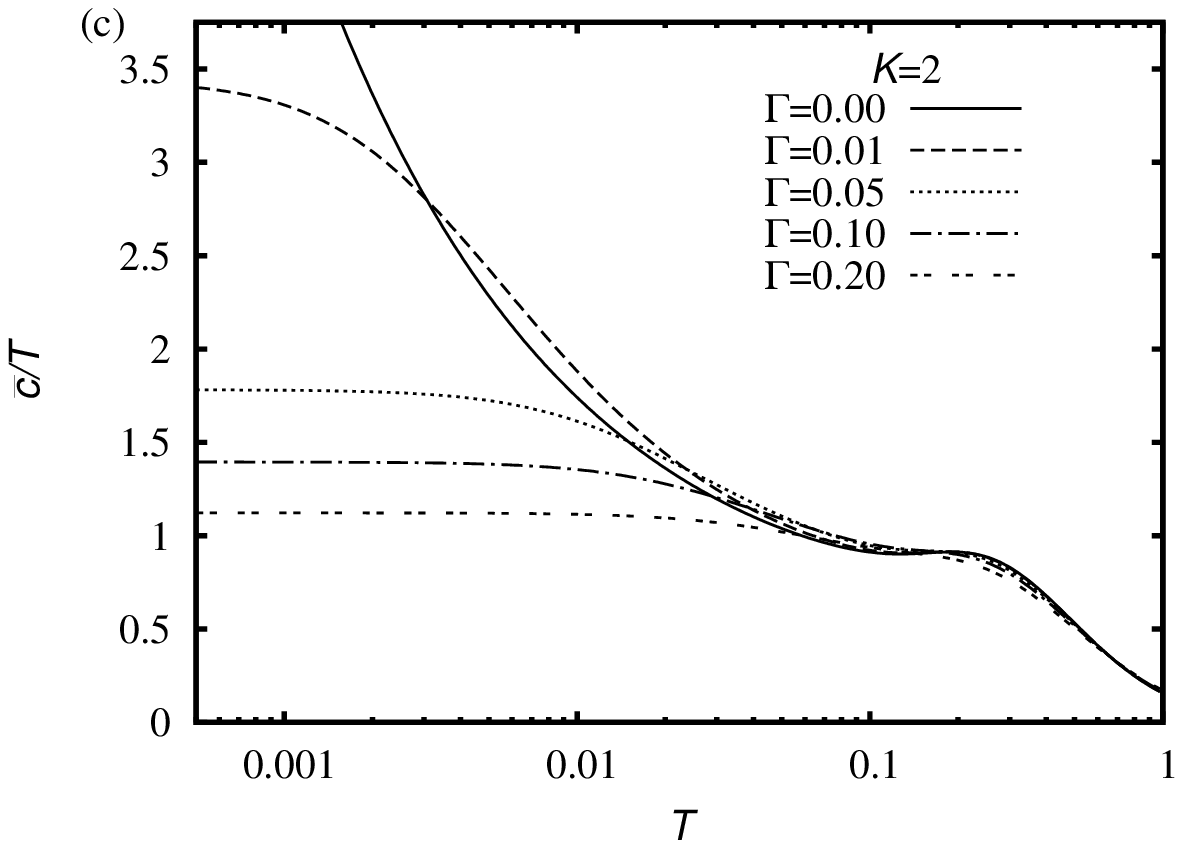}
\includegraphics[clip=on,width=7.5cm,angle=0]{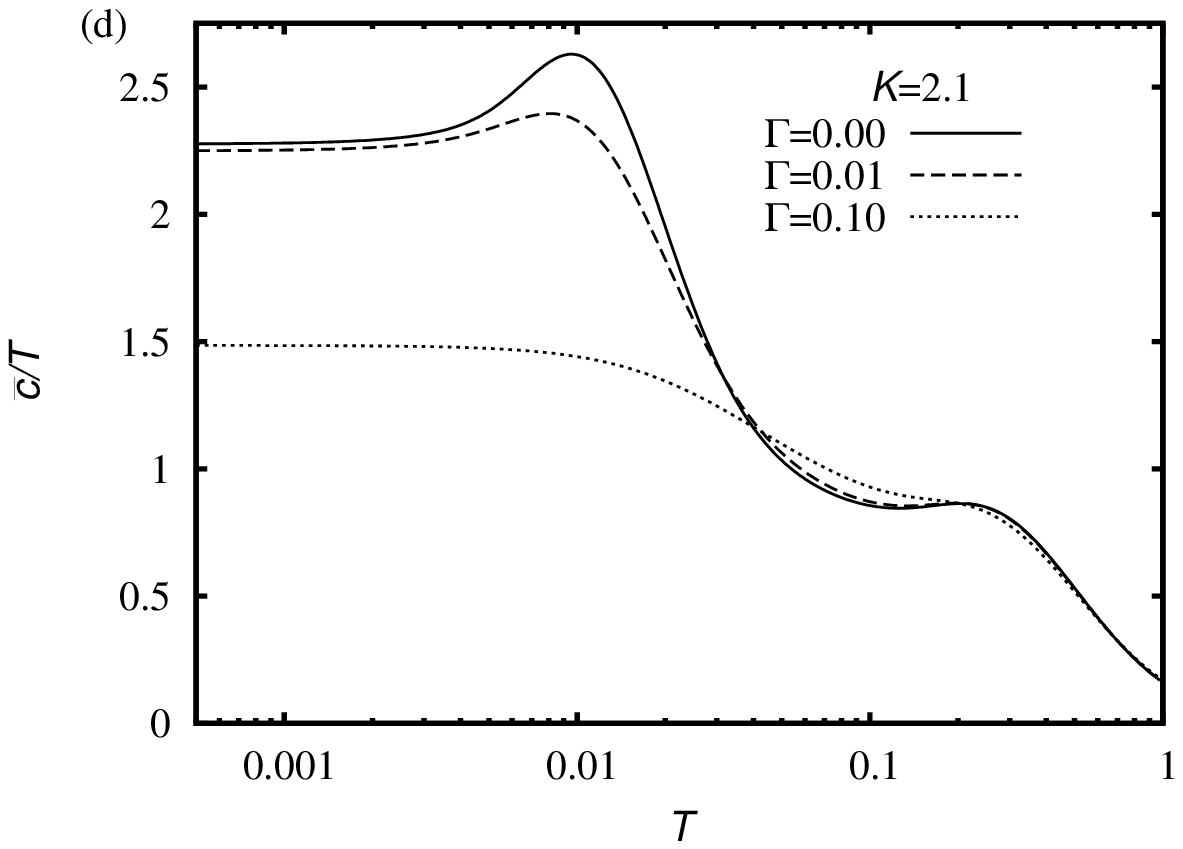}
\caption{$\overline{c}/T$ versus $T$, $\overline{c}$ is the specific heat given in Eq. (\ref{3.03}),
for the spin model (\ref{2.01}), (\ref{2.02}) with $J=1$,
$K=1,\;1.9,\;2,\;2.1$ (from top to bottom),
$\Omega_0=0$,
and
$\Gamma=0,\ldots,0.2$.}
\label{fig04}
\end{figure}

\begin{figure}
\includegraphics[clip=on,width=7.5cm,angle=0]{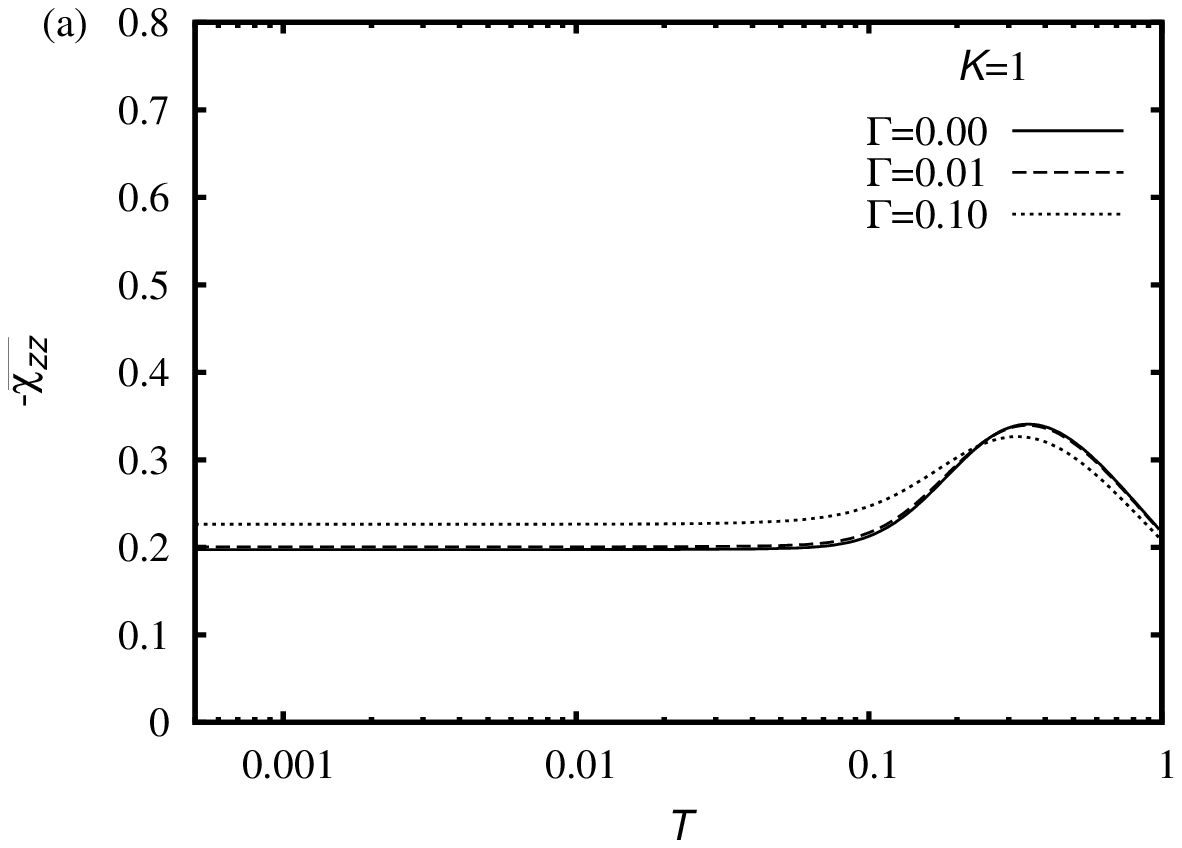}
\includegraphics[clip=on,width=7.5cm,angle=0]{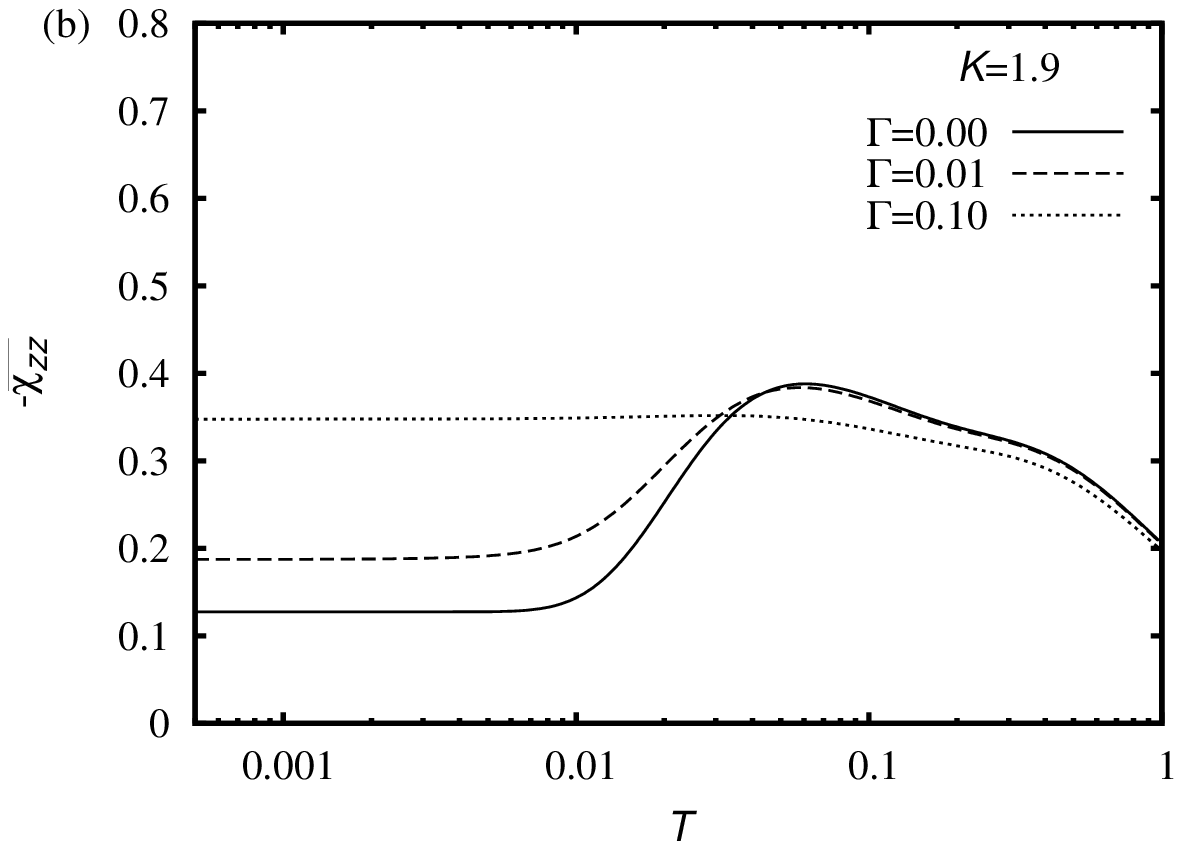}
\includegraphics[clip=on,width=7.5cm,angle=0]{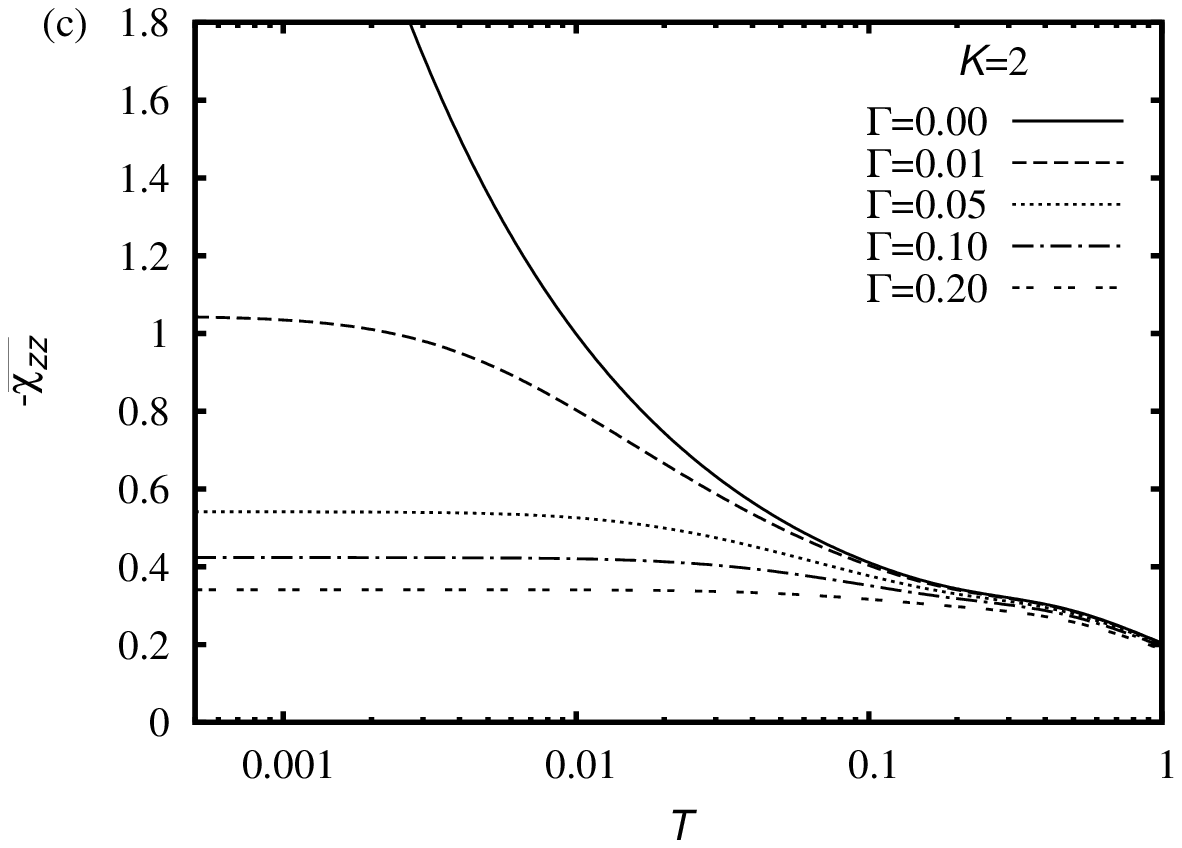}
\includegraphics[clip=on,width=7.5cm,angle=0]{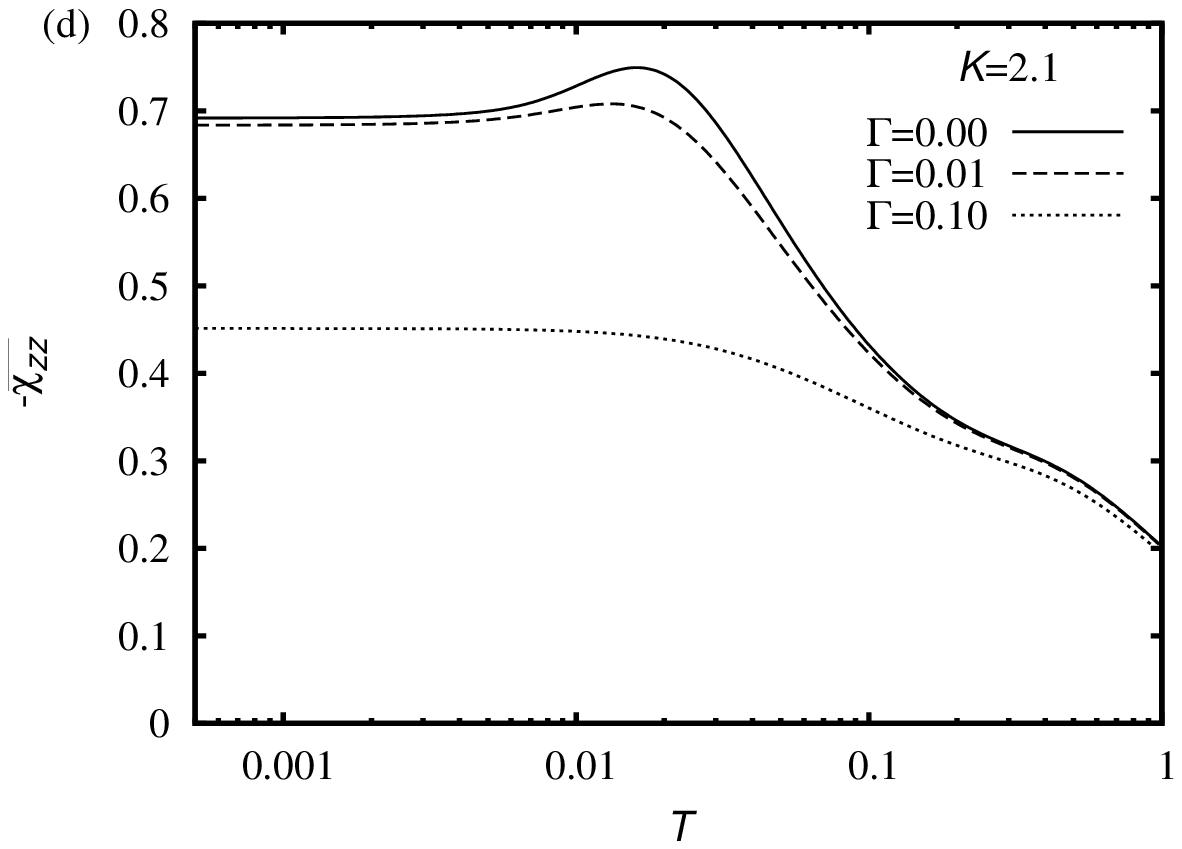}
\caption{Static transverse susceptibility $-\overline{\chi_{zz}}$ (\ref{3.05}) versus $T$
for the spin model (\ref{2.01}), (\ref{2.02}) with $J=1$,
$K=1,\;1.9,\;2,\;2.1$ (from top to bottom),
$\Omega_0=0$,
and
$\Gamma=0,\ldots,0.2$.}
\label{fig05}
\end{figure}

In Figs.~\ref{fig04} and \ref{fig05} we report some of our findings for nonzero temperatures
(for the sake of brevity we consider the case $\Omega_0=0$, only).
The specific heat $\overline{c}$ for various parameter sets is presented in Fig.~\ref{fig04}, 
where we show $\overline{c}/T$ as a function of temperature $T$.
For the nonrandom case $\Gamma=0$ it is known\cite{titvinidze} 
that $c(T)\propto T$ in the limit $T\to 0$ for all $K$ except $K=K_{\rm{crit}}=\pm 2\vert J\vert$.
For the critical value of $K$, $K=K_{\rm{crit}}$, we have $c(T)\propto \sqrt{T}$.

Let us start with the discussion of the behavior of $\overline{c}/T$ for $K=1$, 
i.e., the system is quite far away from the quantum critical point $K_{\rm{crit}}=2$.
Then we have the typical high-temperature maximum in $\overline{c}(T)$ related to the relevant energy scale 
(depending on $J$ and $K$)
and a constant value of $\overline{c}/T$ at low $T$ corresponding to $\overline{c} \propto T$. 
The position of the high-temperature maximum moves to higher temperatures as $K$ increases.
The overall modification of the $\overline{c}/T$ versus $T$ curve by randomness is small. 
There is only a small change of the slope in the dependence $\overline{c}$ on $T$ as $T\to 0$
and a slight shift of the height and the position of the high-temperature maximum.  
At the quantum critical point $K=K_{\rm{crit}}=2$ we have a completely different behavior of $\overline{c}/T$. 
The high-temperature maximum is still present 
(and again there is only a weak effect of randomness on that maximum). 
However, below the maximum there is an increase of $\overline{c}/T$ with decreasing $T$ 
indicating the $\overline{c} \propto \sqrt{T}$ dependence.
While this increase is monotonous till $T \to 0$ for the nonrandom case,
in the random system there is only a finite region of $T$ 
where this increase of $\overline{c}/T$ can be observed. 
At lower $T$ again the $\overline{c} \propto T$ regime sets in, 
i.e., the randomness destroys the $\sqrt{T}$-dependence in favor of the $T$-dependence as $T\to 0$. 
A similar behavior can be found for $K$ values near $K_{\rm{crit}}=2$,
e.g., at $K=1.9$ or $2.1$, 
i.e., we observe the $\sqrt{T}$-dependence 
for the low-temperature specific heat in a finite temperature range below the high-temperature maximum. 
That is the typical quantum critical behavior 
appearing in the vicinity of a quantum critical point.\cite{qpt1,qpt2,qpt2a}
Interestingly, 
in the specific model under consideration for small randomness 
there is a second maximum for $\overline{c}/T$ 
(but not for $\overline{c}$) 
at a lower temperature $T^{\star}$ 
(which is a reminiscence of the singularity for $K=K_{\rm{crit}}$ at $T=0$) 
before the $\overline{c} \propto T$ regime sets in at very low temperatures.  
According to the above discussion for Figs.~\ref{fig04}b, \ref{fig04}c, and \ref{fig04}d, 
we conclude that for the system at $K=K_{\rm{crit}}$ at low but finite temperatures 
the randomness has a similar effect as shifting the nonrandom system slightly away from the quantum critical point.
Note that further increasing of $\Gamma$ removes the low-temperature maximum in  $\overline{c}/T$ 
and also the  critical-like $\sqrt{T}$ behavior of $\overline{c}$ disappears,
see the curves for $\Gamma=0.1$ in panels for $K=1.9,\;2.1$ of Fig.~\ref{fig04}.

The temperature dependence of the static transverse susceptibility $\overline{\chi_{zz}}$ shown in Fig.~\ref{fig05}
exhibits many similarities to that of $\overline{c}/T$ discussed above.
In the nonrandom case it is known\cite{titvinidze} 
that for $K=K_{\rm{crit}}$ one has $\chi_{zz}\propto 1/\sqrt{T}$ (critical behavior) as $T\to 0$, 
whereas $\chi_{zz}$ remains finite at $T=0$ for noncritical values of $K$.
For $K$ around $K_{\rm{crit}}$ a reminiscent of the critical behavior emerges in the low-temperature region 
starting from certain finite temperatures.
For small nonzero $\Gamma$, 
$\overline{\chi_{zz}}$ may exhibit the critical behavior in a certain temperature range starting from finite temperatures 
if $K$ is equal to or is close to $K_{\rm{crit}}$,
see the curves for $\Gamma=0.01$ in panels $K=1.9,\;2,\;2.1$ of Fig.~\ref{fig05}.

\begin{figure}
\includegraphics[clip=on,width=7.5cm,angle=0]{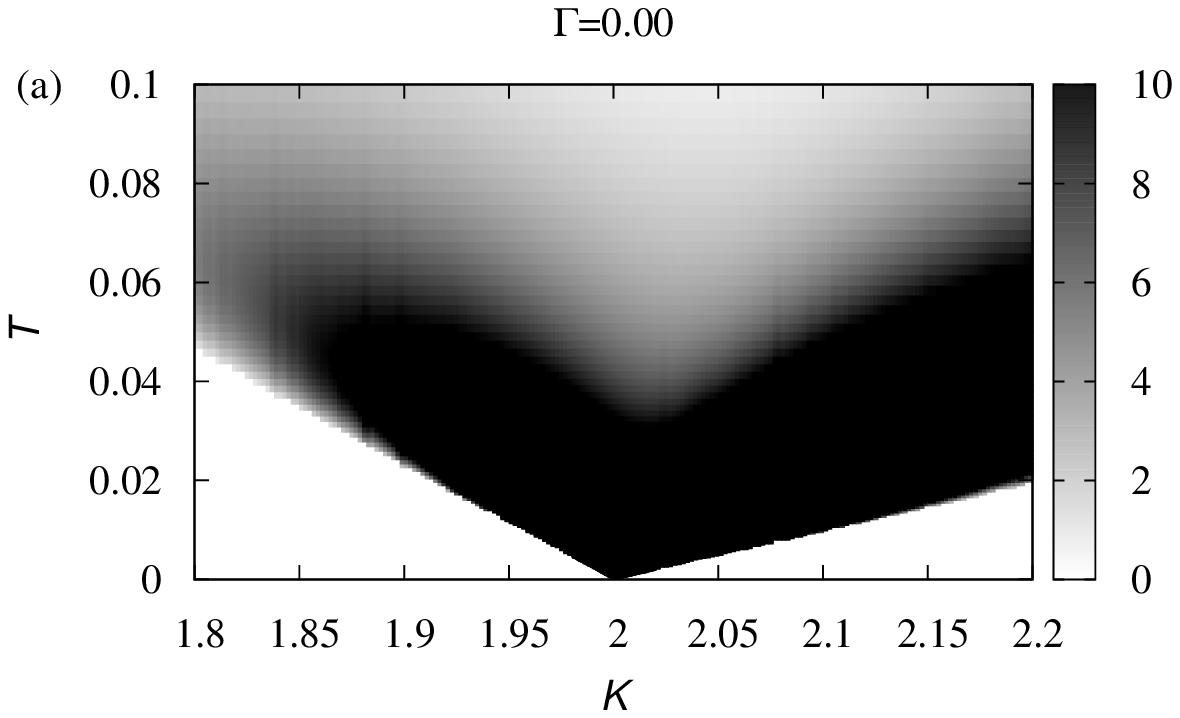}
\includegraphics[clip=on,width=7.5cm,angle=0]{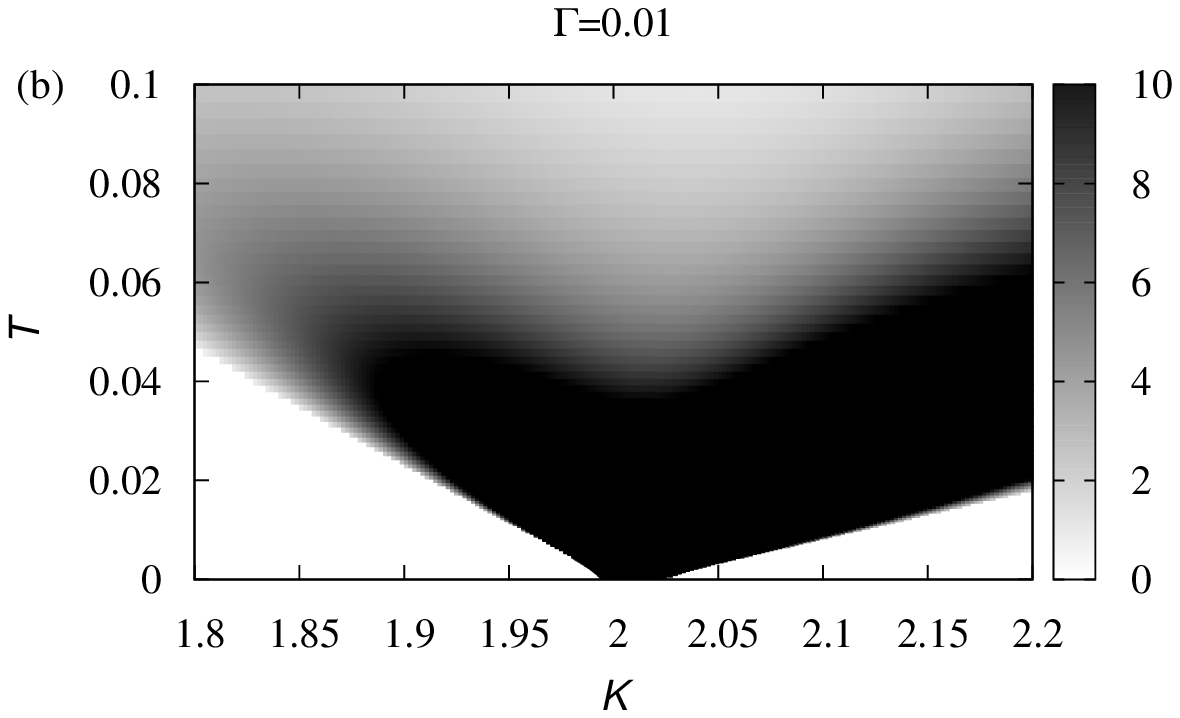}
\includegraphics[clip=on,width=7.5cm,angle=0]{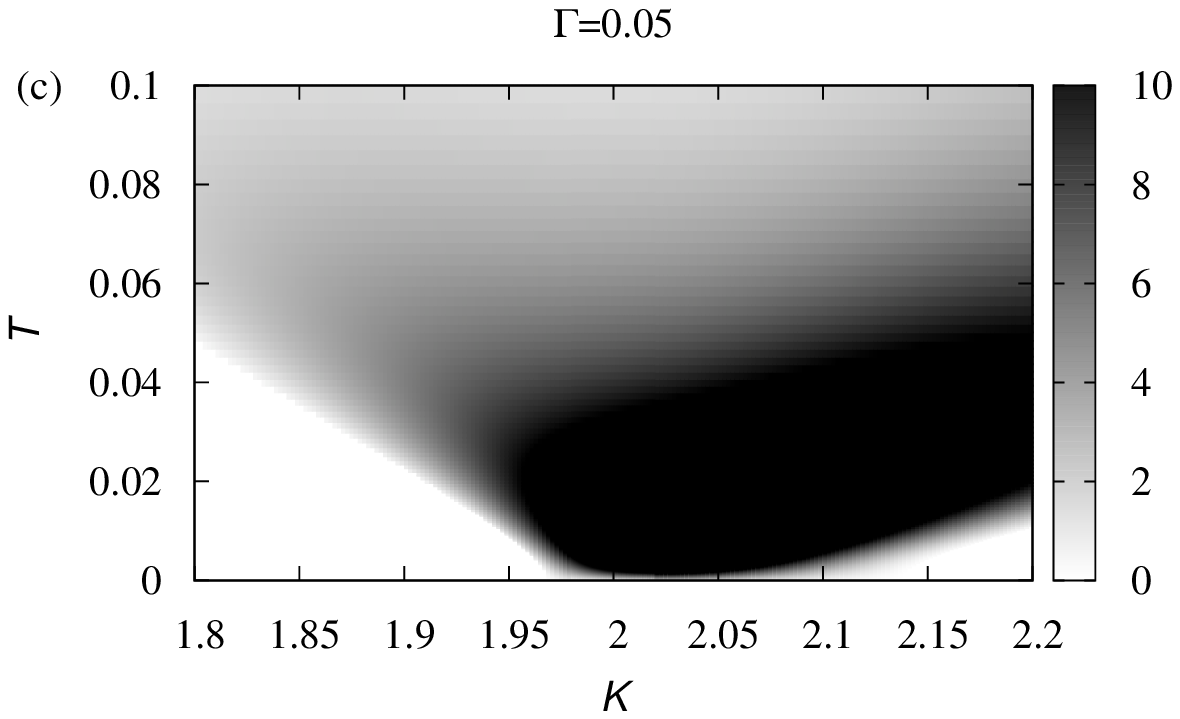}
\includegraphics[clip=on,width=7.5cm,angle=0]{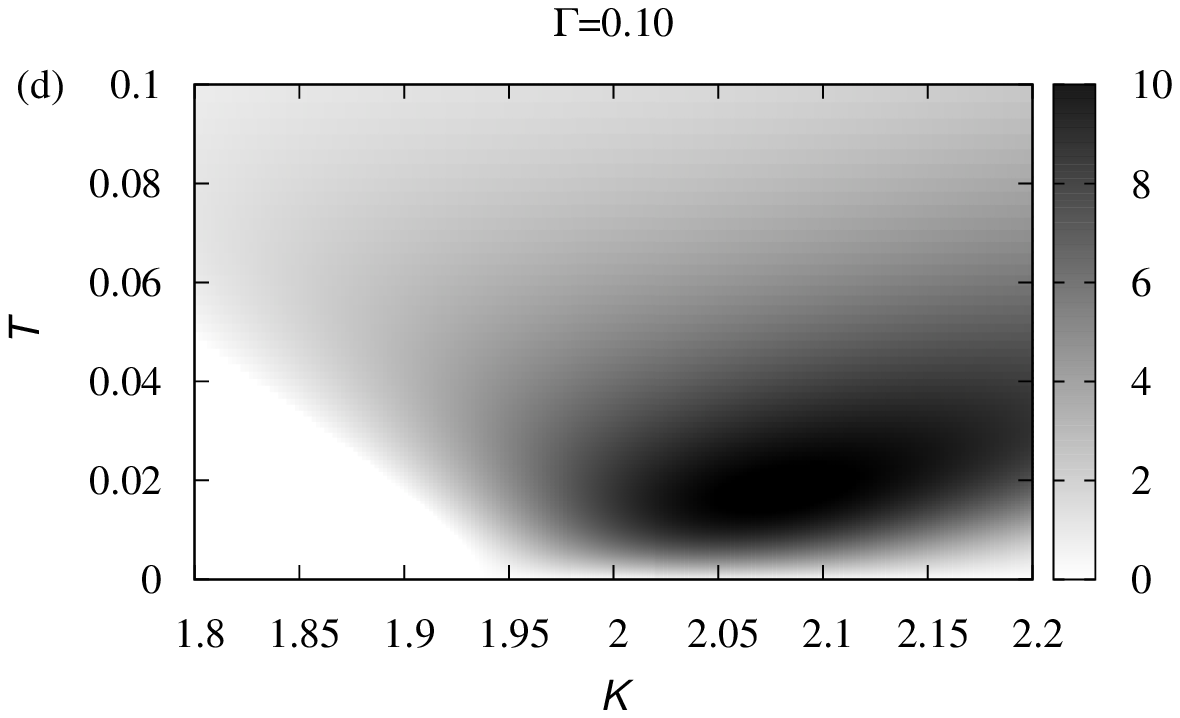}
\caption{Critical region as indicated by the value of $-\partial(\overline{c}/T)/\partial T>0$ 
in the half-plane $K$--$T$
for the spin model (\ref{2.01}), (\ref{2.02}) with $J=1$, $\Omega_0=0$ and $\Gamma=0,\;0.01,\;0.05,\;0.1$ 
(from top to bottom).
See further explanations in the main text.}
\label{fig06}
\end{figure}
\begin{figure}
\includegraphics[clip=on,width=7.5cm,angle=0]{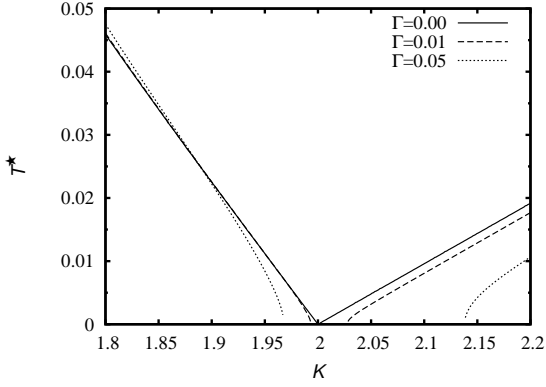}
\caption{The lower boundary of the quantum critical region
determined by the position of the maximum $T^{\star}$ of  $\overline{c}/T$
for the spin model (\ref{2.01}), (\ref{2.02}) with $J=1$, $\Omega_0=0$ and $\Gamma=0,\;0.01,\;0.05$.}
\label{fig07}
\end{figure}

We can use the above discussed observations of the temperature profiles of $\overline{c}/T$ and $\overline{\chi_{zz}}$
to construct a phase diagram of the random quantum spin chain (\ref{2.01}), (\ref{2.02}), 
i.e., to determine the quantum critical region in the  $K$--$T$ half-plane, see Fig.~\ref{fig06}.
As an indicator of the critical (i.e., $\sqrt{T}$-like) behavior 
we choose the value of the derivative $-\partial(\overline{c}/T)/\partial T$.
To be more specific,
we use the circumstance that for $K$ in the vicinity of $K_{{\rm{crit}}}$ and small $\Gamma\ge 0$
$\overline{c}/T$ exhibits a maximum at $T^\star$
(for $K=K_{{\rm{crit}}}$, $\Gamma= 0$ it is a divergency at $T=0$)
and a $1/\sqrt{T}$-like decrease
(due to $\overline{c} \propto  \sqrt{T}$)
in a certain temperature region above $T^\star$.
As discussed above this behavior is a trace of the quantum critical point.
The plots shown in Fig.~\ref{fig06} 
are based on a quantitative analysis of $-\partial(\overline{c}/T)/\partial T$ 
and show the value of $-\partial(\overline{c}/T)/\partial T$ as grayscale plots. 
All white areas in this figure belong to negative values of $-\partial(\overline{c}/T)/\partial T$. 
The lower boundary of the quantum critical region is related to $T^\star$ 
(the temperature where the low-temperature maximum of $\overline{c}/T$ is located),
since for $T<T^\star$ one has $-\partial(\overline{c}/T)/\partial T < 0$, 
whereas already for $T$ slightly above $T^\star$ 
the derivative $-\partial(\overline{c}/T)/\partial T >0$ becomes quite large. 
As a result, the lower boundary of the quantum critical region is quite sharp in Fig.~\ref{fig06}.
Note that for $\Gamma=0$ our plot reproduces the typical picture for a quantum critical region.\cite{qpt1,qpt2,qpt2a}
For small $\Gamma\ge 0$ the lower boundaries in the $K$-$T$ half-plane remain nearly straight lines 
with different slopes $\approx-0.225$ and $\approx 0.096$ below and above $K=2$, respectively.
We compare these lower boundaries for various strengths of disorder $\Gamma$ in Fig.~\ref{fig07}, 
where the numerically determined values of $T^{\star}$ as a function of $K$ are shown.
It is obvious that a noticeable change in the slope is observed only in a small vicinity of the $K=2$.
On the other hand, 
the upper boundary of the quantum critical region is not sharp, 
since $-\partial(\overline{c}/T)/\partial T$ varies smoothly, 
if the temperature is further growing.
From Fig.~\ref{fig06} it is obvious that for small randomness 
(see the panel for $\Gamma=0.01$ in Fig.~\ref{fig06} and the corresponding curves in Fig.~\ref{fig07})
at low temperatures the area, where signatures of quantum critical behavior can be observed, is enlarged, 
whereas a further increase of randomness then leads to a shrinking of that area. 
For $\Gamma \gtrsim 0.1$ the signatures of quantum criticality completely disappear.

We have to say in the end,
that the adopted criterion to indicate the effect of randomness on quantum critical behavior 
represents only one possibility.
In fact, 
with this criterion 
we only indicate a region where the specific heat $\overline{c}$ behaves like $T^{\alpha}$ with $\alpha<1$ 
and claim that such a behavior is a remnant of the quantum critical one in the nonrandom model when $\alpha=1/2$.

\section{Generalization of the model}
\label{sec4}
\setcounter{equation}{0}

Our consideration until now was restricted to the $XX$ two-site interaction and the $XZX+YZY$ three-site interaction 
[see Eq. (\ref{2.01})] 
with a Lorentzian probability distribution of the random transverse field, 
which allows the exact calculation of averaged density of states.
In this section we want to generalize the model 
by (i) extending the Hamiltonian and (ii) considering general probability distributions.

First we illustrate the extension of the Hamiltonian. 
We can add to the Hamiltonian (\ref{2.01}) 
a Dzyaloshinskii-Moriya two-site interaction with the strength $D_n$
and 
a $XZY-YZX$ three-site interaction with the strength $E_n$,
\begin{eqnarray}
\label{4.01}
H=\sum_{n}
\left[J_n\left(s_n^xs_{n+1}^x+s_n^ys_{n+1}^y\right)
\right.
\nonumber\\
\left.
+D_n\left(s_n^xs_{n+1}^y-s_n^ys_{n+1}^x\right)
\right.
\nonumber\\
\left.
+K_n\left(s_n^xs^z_{n+1}s_{n+2}^x+s_n^ys^z_{n+1}s_{n+2}^y\right)
\right.
\nonumber\\
\left.
+E_n\left(s_n^xs^z_{n+1}s_{n+2}^y-s_n^ys^z_{n+1}s_{n+2}^x\right)\right]
+\sum_{n}\Omega_ns_n^z.
\end{eqnarray}
In fermionic representation Eq. (\ref{4.01}) reads
\begin{eqnarray}
\label{4.02}
H=\sum_{n}
\left[{\cal{A}}_n\left(c_n^{\dagger}c_n-\frac{1}{2}\right)
+\left({\cal{B}}_nc_n^{\dagger}c_{n+1}+{\rm{H.c.}}\right)
\right.
\nonumber\\
\left.
+\left({\cal{C}}_nc_n^{\dagger}c_{n+2}+{\rm{H.c.}}\right)\right],
\end{eqnarray}
where ${\cal{A}}_n=\Omega_n$,
${\cal{B}}_n=(J_n+{\rm{i}}D_n)/2$,
and
${\cal{C}}_n=-(K_n+{\rm{i}}E_n)/4$.
Comparing 
Eq. (\ref{4.02}) with Eq.~(\ref{2.03})
we conclude 
that for a Lorentzian probability of $\Omega_n$ 
and uniform parameters $J_n=J$, $D_n=D$, $K_n=K$, and $E_n=E$ 
the formula (\ref{2.11}) for the averaged Green function is valid also for the generalized model (\ref{4.01}) 
if we change   
$J\cos\kappa-(K/2)\cos(2\kappa)
\to
J\cos\kappa+D\sin\kappa-(K/2)\cos(2\kappa)-(E/2)\sin(2\kappa)$.
Further analysis can be performed as in the previous sections.
Although the results obtained from the exact averaged density of states are more general now, 
they do not give basically new features in comparison to those discussed above.

So far our discussion has been referred to a special kind of randomness,
i.e., to a Lorentzian transverse field.
Only in this case we can provide an exact analysis of the thermodynamics of the spin model.
However, it is possible to analyze  some global properties of the averaged density of states 
(which imply corresponding properties of the spin chain)
for an arbitrary inhomogeneous spin-1/2 $XX$ chain with three-site interactions in a transverse field (\ref{4.01}).
For that we consider the moments of the density of states (\ref{2.06})\cite{elk}
\begin{eqnarray}
\label{4.03}
M^{(0)}&=&\int{\rm{d}}\omega\rho(\omega)
=\frac{1}{N}\sum_{n=1}^N\left\langle\left\{c_n,c_n^{\dagger}\right\}\right\rangle=1,
\nonumber\\
M^{(1)}&=&\int{\rm{d}}\omega \omega \rho(\omega)
=\frac{1}{N}\sum_{n=1}^N\left\langle\left\{\left[c_n,H\right],c_n^{\dagger}\right\}\right\rangle,
\nonumber\\
M^{(2)}&=&\int{\rm{d}}\omega \omega^2 \rho(\omega)
=\frac{1}{N}\sum_{n=1}^N\left\langle\left\{\left[\left[c_n,H\right],H\right],c_n^{\dagger}\right\}\right\rangle,
\nonumber\\
M^{(3)}&=&\int{\rm{d}}\omega \omega^3 \rho(\omega)
=\frac{1}{N}\sum_{n=1}^N
\left\langle\left\{\left[\left[\left[c_n,H\right],H\right],H\right],c_n^{\dagger}\right\}\right\rangle
\nonumber\\
\end{eqnarray}
etc.
Calculating the right-hand sides in Eq. (\ref{4.03}) with the Hamiltonian (\ref{4.02}) 
we find
\begin{eqnarray}
\label{4.04}
&&M^{(1)}=\frac{1}{N}\sum_n{\cal{A}}_n,
\nonumber\\
&&M^{(2)}=\frac{1}{N}\sum_n\left({\cal{A}}_n^2
+\vert{\cal{C}}_{n-2}\vert^2 + \vert{\cal{B}}_{n-1}\vert^2
+\vert{\cal{B}}_{n}\vert^2 + \vert{\cal{C}}_{n}\vert^2\right),
\nonumber\\
&&M^{(3)}=\frac{1}{N}\sum_n\left[\vert{\cal{C}}_{n-2}\vert^2{\cal{A}}_{n-2}
+\vert{\cal{B}}_{n-1}\vert^2{\cal{A}}_{n-1}
\right.
\nonumber\\
&&\left.
+{\cal{A}}_{n}^3
+2\left(\vert{\cal{C}}_{n-2}\vert^2+\vert{\cal{B}}_{n-1}\vert^2
+\vert{\cal{B}}_{n}\vert^2+\vert{\cal{C}}_{n}\vert^2\right){\cal{A}}_{n}
\right.
\nonumber\\
&&\left.
+\vert{\cal{B}}_{n}\vert^2{\cal{A}}_{n+1}+\vert{\cal{C}}_{n}\vert^2{\cal{A}}_{n+2}
\right.
\nonumber\\
&&\left.
+2\Re \left({\cal{C}}^{*}_{n-2}{\cal{B}}_{n-2}{\cal{B}}_{n-1}
+{\cal{C}}^{*}_{n-1}{\cal{B}}_{n-1}{\cal{B}}_{n}
+{\cal{C}}^{*}_{n}{\cal{B}}_{n}{\cal{B}}_{n+1}\right)
\right].
\nonumber\\
\end{eqnarray}
The moments of the random-averaged density of states $\overline{\rho(\omega)}$ 
follow from Eq. (\ref{4.04}) after a corresponding averaging.

We may use the derived moments of the density of states (\ref{4.04}) 
to examine some general properties of the (homogeneous or inhomogeneous) spin chain (\ref{4.01}).
For example, the ground-state transverse magnetization is given by the formula
\begin{eqnarray}
\label{4.05}
m_z=-\int{\rm{d}}\omega\rho(\omega)\left[\theta(\omega)-\frac{1}{2}\right],
\end{eqnarray}
see Eq. (\ref{3.04}).
It has been shown that the uniform spin model (\ref{4.01}) may exhibit a nonzero transverse magnetization $m_z$ 
in zero transverse field $\Omega_n=\Omega_0=0$.\cite{gottlieb_drd,titvinidze,lou,footnote1}
From Eq. (\ref{4.05}) it is clear 
that $m_z\ne 0$ at zero field if the density of states is asymmetric, 
i.e., if the third moment of the density of states at zero field is nonzero.

In the uniform (nonrandom) case we have
$M^{(3)}={\cal{A}}^3+6(\vert{\cal{B}}\vert^2+\vert{\cal{C}}\vert^2){\cal{A}}+6\Re({\cal{C}}^*{\cal{B}}^2)$,
or in the zero-field case
$M^{(3)}=6\Re({\cal{C}}^*{\cal{B}}^2)=-(3/8)[(J^2-D^2)K+2JDE]$.
From the latter formula it is obvious 
that only three-site interactions may lead to nonzero magnetization.
More precisely, 
the $XZX+YZY$ interaction if $J^2\ne D^2$ 
or
the $XZY-YZX$ interaction if $JD\ne 0$
leads to $m_z\ne 0$ at zero field.
Formulas for the moments of the density of states (\ref{4.04}) 
permit to examine the effect of randomness on this property.
An example of such analysis is given in Ref.~\onlinecite{gottlieb_drd}b
where some consequences of the correlated off-diagonal and diagonal disorder were discussed. 

Finally we note
that the high-temperature properties of the spin model are determined by the lower moments of the density of states
[see Eq. (\ref{2.06})] 
and therefore the thermodynamic quantities in the high-temperature limit 
can be examined accurately for any type of disorder on the basis of Eq. (\ref{4.04}).

\section{Conclusions}
\label{sec5}
\setcounter{equation}{0}

We have considered a random spin-1/2 $XX$ chain with three-site interactions.
For random on-site transverse field with Lorentzian probability distribution
we have calculated exactly the random-averaged density of states 
and the corresponding random-averaged thermodynamic quantities of the model.
For arbitrary inhomogeneous Hamiltonian parameters we have calculated the 
three first moments of the density of states 
which determine some general properties of the spin model 
and yield its thermodynamic quantities in the high-temperature limit.
The effect of a random transverse field on the ground-state magnetization process 
and on the temperature behavior of the specific heat and static transverse susceptibility
has been analyzed.
As a main result 
we have discussed how the quantum critical behavior of the spin-1/2 $XX$ chain with three-site interactions
is modified by randomness.
While for large enough randomness all signatures of quantum critical behavior disappear, 
we find even a slightly enlarged temperature area for very small randomness, 
where such signatures can be observed.

Finally, we have to underline 
that our analytical findings are restricted to the random-averaged density of states and thermodynamic quantities.
To our best knowledge,
almost all applications of Lloyd's model are restricted to one-particle quantities which can be obtained rigorously.
The calculation of the two-particle quantities meets notorious difficulties and requires approximations,
see, e.g., Ref.~\onlinecite{saitoh}.
On the other hand,
other quantities of interest 
which are not related to the one-particle Green functions (\ref{2.08})
but are important for following final outcomes of introduced randomness
(e.g., two-spin correlation functions)
may be examined only numerically for finite chains,\cite{numerics}
see also Ref.~\onlinecite{joachim}.
However, numerical studies are beyond the scope of the present paper and will be reported separately.

\section*{Acknowledgments}

The authors thank J.~J\c{e}drzejewski for discussions.
O.~D. acknowledges the financial support of the DAAD and thanks Magdeburg University for hospitality in 2007-8 and in 2010.
J.~R. acknowledges the financial support by DFG (project RI615/16-1).
The authors thank the anonymous referee for a number of useful comments and remarks.

\appendix
\section{The roots of Eq. (\ref{2.13})}
\label{a}

In this appendix we give explicit expressions for the roots of Eq. (\ref{2.13}).
They read:
\begin{eqnarray}
\label{A.1}
\left\{z_1,z_2,z_3,z_4\right\}
=\left\{
\frac{\frac{J}{K}-g-\sqrt{\left(\frac{J}{K}-g\right)^2-4}}{2},
\right.
\nonumber\\
\left.
\frac{\frac{J}{K}+g-\sqrt{\left(\frac{J}{K}+g\right)^2-4}}{2},
\right.
\nonumber\\
\left.
\frac{\frac{J}{K}-g+\sqrt{\left(\frac{J}{K}-g\right)^2-4}}{2},
\right.
\nonumber\\
\left.
\frac{\frac{J}{K}+g+\sqrt{\left(\frac{J}{K}+g\right)^2-4}}{2}
\right\}
\end{eqnarray}
with
$g=\sqrt{(J/K)^2-(4/K)(\omega-\Omega_0+{\rm{i}}\Gamma)+2}$
in the case of the retarded Green functions (\ref{2.08})
or 
$g=\sqrt{(J/K)^2-(4/K)(\omega-\Omega_0-{\rm{i}}\Gamma)+2}$
in the case of the advanced Green functions (\ref{2.08}).


\begin{thebibliography}{99}
\bibitem{footnote1}
For (nonrandom) spin-1/2 $XX$ chains with the $XZY-YZX$ three-site interactions 
it is immediately clear that $m_z= 0$ if $\Omega=0$.
Really, 
consider a spin rotation around the $y$ axis by $\pi$,
$s^y \to  s^y$,
$s^x \to -s^x$,
$s^z \to -s^z$.
As a result,
the Hamiltonian $H(J, E, \Omega)$ transforms into the Hamiltonian $H(J, E, -\Omega)$,
and for $\Omega=0$ we have 
$\langle s^z\rangle_{J,E,\Omega=0}=-\langle s^z\rangle_{J,E,-\Omega=0}=0$.
\end{thebibliography}
\end{document}